\documentclass[12pt]{article}
\pdfoutput=1
\usepackage[letterpaper, margin=1in]{geometry}
\usepackage{amsmath, amssymb}
\usepackage{cite}
\usepackage{graphicx}
\usepackage[font=small,labelfont=bf]{caption}
\usepackage[normalem]{ulem}

\usepackage{xcolor}
\usepackage{slashed}
\usepackage{cancel}
\usepackage{bm}
\usepackage{verbatim}
\usepackage{tabularx}

\definecolor{DarkRed}{rgb}{0.6,0,0}
\definecolor{DarkGreen}{rgb}{0,0.6,0}
\definecolor{DarkBlue}{rgb}{0,0,0.6}
\definecolor{DarkOrange}{rgb}{1,0.35,0}
\definecolor{PouyaGreen}{rgb}{0,0.66,0}
\definecolor{PouyaPurple}{rgb}{0.5,0,0.5}
\definecolor{matBlue}{HTML}{1F77B4}
\definecolor{matOrange}{HTML}{FF7F0E}
\definecolor{matGreen}{HTML}{2CA02C}
\definecolor{matRed}{HTML}{D62728}
\definecolor{goodgreen}{rgb}{0,.6,0.4}
\definecolor{UOgreen}{HTML}{007030}

\usepackage{xkcdcolors}
\usepackage{setspace}
\usepackage{authblk}
\usepackage{physics}
\usepackage{multirow}

\usepackage{bbm}

\usepackage{dsfont}

\usepackage{etoolbox}

\let\OLDthebibliography\thebibliography
\renewcommand\thebibliography[1]{
	\OLDthebibliography{#1}
	\setlength{\parskip}{3.0pt plus 2.5pt minus 1.0pt}
	\setlength{\itemsep}{3.0pt plus 2.5pt minus 1.0pt}
}

\usepackage[greek,english]{babel}

\usepackage[colorlinks=true,linktocpage=true,linkcolor=DarkOrange,citecolor=UOgreen,urlcolor=UOgreen]{hyperref}

\usepackage{cleveref}
\crefname{table}{Table}{Tables}
\crefname{equation}{Eq.}{Eqs.}
\crefname{eqnarray}{Eq.}{Eqs.}
\crefname{appendix}{App.}{Apps.}
\crefname{section}{Sec.}{Secs.}
\crefname{figure}{Fig.}{Figs.}

\graphicspath{{figs/}}

\begin{document}

\title{\bf Fermion-Portal Dark Matter at a High-Energy Muon Collider}

\author{Pouya Asadi$^1$, Samuel Homiller$^2$, Aria Radick$^1$, Tien-Tien Yu$^1$\\
{\small \color{UOgreen} \texttt{pasadi@uoregon.edu},~ \texttt{shomiller@cornell.edu},~ \texttt{aradick@uoregon.edu},~\texttt{tientien@uoregon.edu}, }\\
{\small${}^1$Institute for Fundamental Science and Department of Physics,}\\
{\small University of Oregon, Eugene, OR 97403}\\
{\small${}^2$Laboratory for Elementary Particle Physics, Cornell University, Ithaca, NY 14853}
}

\date{\today}

\maketitle


\begin{abstract}
In this work, we provide a comprehensive study of fermion-portal dark matter models in the freeze-in regime at a future muon collider. For different possible non-singlet fermion portals, we calculate the upper bound on the mediator's mass arising from the relic abundance calculation and discuss the reach of a future muon collider in probing their viable parameter space in prompt and long-lived particle search strategies. 
In particular, we develop rudimentary search strategies in the prompt region and show that cuts on the invariant dilepton or dijet masses, the missing transverse mass $M_{T2}$, pseudorapidity and energy of leptons or jets, and the opening angle between the lepton or the jet pair can be employed to subtract the Standard Model background. 
In the long-lived particle regime, we discuss the signals of each model and calculate their event counts. 
In this region, the lepton-(quark-)portal model signal consists of charged tracks ($R$-hadrons) that either decay in the detector to give rise to a displaced lepton (jet) signature, or are detector stable and give rise to heavy stable charged track signals. 
As a byproduct, a pipeline is developed for including the non-trivial parton distribution function of a muon component inside a muon beam; it is shown that this leads to non-trivial effects on the kinematic distributions and attainable significances. 
We also highlight phenomenological features of all models unique to a muon collider and hope our results, for this motivated and broad class of dark matter models, inform the design of a future muon collider detector. We also speculate on suggestions for improving the sensitivity of a muon collider detector to long-lived particle signals in fermion-portal models.
\end{abstract}

\newpage

\tableofcontents

\begin{spacing}{1.2}
\section{Introduction}
\label{sec:intro}
The only concrete evidence we have for dark matter (DM) thus far comes from its gravitational effects. Understanding the particle nature of DM, such as its mass, spin, and non-gravitational interactions, remains a major goal in particle physics, spurring extensive experimental and theoretical research (see, \textit{e.g.},~Ref.\cite{Cirelli:2024ssz} for a recent review). Despite these efforts, we have yet to observe any clear signs of DM's non-gravitational interactions, suggesting that any such interactions with the Standard Model (SM) must be minimal. One approach to systematically study potential interactions considers DM residing in a hidden sector, with its only connections to the SM mediated through a finite number of renormalizable ``portal" interactions~\cite{Pospelov:2007mp}. As a result, the phenomenology of the DM is encapsulated in the details of the portal interactions, specifically the mediator and its SM interactions, rather than those of the hidden sector. 

One such portal is the ``fermion portal"~\cite{Bai:2013iqa,Bai:2014osa}, which is the focus of this work. Here, the mediator particle has the same quantum numbers as one of the SM fermions, allowing for a gauge-singlet operator and thus a connection between the DM candidate and the SM. This class of models has also been studied under other names, such as flavored DM \cite{Agrawal:2011ze,Batell:2013zwa,Agrawal:2014una,Agrawal:2014aoa,Agrawal:2015tfa,Agrawal:2015kje,Agrawal:2016uwf,Desai:2020rwz,Acaroglu:2022hrm}, effective WIMPs~\cite{Chang:2013oia,Chang:2014tea}, or color-charged mediator models \cite{DiFranzo:2013vra,Garny:2014waa}. These models give rise to numerous interesting experimental signals \cite{Bai:2013iqa,Bai:2014osa,Agrawal:2011ze,Batell:2013zwa,Agrawal:2014una,Agrawal:2014aoa,Agrawal:2015tfa,Agrawal:2015kje,Agrawal:2016uwf,Desai:2020rwz,Acaroglu:2022hrm,Chang:2013oia,Chang:2014tea,DiFranzo:2013vra,Garny:2014waa,Evans:2016zau,Garny:2018ali,Calibbi:2021fld,DAmbrosio:2021wpd}, foremost among them a host of signals at colliders. 
While this setup allows us to have multiple flavors of DM particle, which can open the doors for many interesting model-building efforts (\textit{e.g.} see Refs.~\cite{Hamze:2014wca,Agrawal:2016uwf,Dessert:2018khu}), the collider signals of the model are for the most part independent of the number of flavors. Instead, the precise phenomenology depends on the specific quantum numbers of the mediator. This includes the mediator production cross sections, its decay channels and branching ratios, as well as the SM backgrounds and final state signatures. Importantly, the production channels and background can vary significantly between hadron and lepton colliders, warranting dedicated studies for different machines. 

Much of the previous literature on fermion--portal DM has focused on hadron colliders, such as the LHC and future colliders such as HL-LHC and a proposed 100 TeV $pp$ collider.
However, muon colliders have been proposed as a viable path forward to higher energies with vast potential for studying new theories beyond the Standard Model (BSM)~\cite{Ankenbrandt:1999cta,Wang_2016,Boscolo:2018ytm,Neuffer:2018yof,Delahaye:2019omf,Shiltsev:2019rfl,MuonCollider:2022xlm,MuonCollider:2022glg,MuonCollider:2022nsa,MuonCollider:2022ded,Black:2022cth,Accettura:2023ked,InternationalMuonCollider:2024jyv}.\footnote{See also Ref.~\cite{Hamada:2022mua} for proposals for $\mu^+\mu^+$ and $\mu^+e^-$ colliders with interesting potential reach in BSM models.} 
Such a machine has recently emerged as the leading proposal for a lepton collider to push the boundaries of the energy frontier. 
The design of such a collider and its detector, which has been referred to as MUSIC,\footnote{
{MUon Smasher for Interesting Collisions}~\cite{MuCoL:2024oxj}. While other detector designs exist, we will specialize to the MUSIC design for a muon collider detector in this work.} is an active field of research.

In this work, we expand the studies of fermion-portal DM at muon colliders. Specifically, we focus on models in which the DM abundance is set through a freeze-in mechanism, which is a consequence of very feeble couplings between the SM and the dark sector, and complements previous work that primarily focused on models with freeze-out production. The freeze-in mechanism of fermion-portal DM has been studied in Refs.~\cite{Evans:2016zau,Garny:2018ali,Calibbi:2021fld,DAmbrosio:2021wpd,Asadi:2023csb}. 
In particular, Refs.~\cite{Garny:2018ali,Asadi:2023csb} argued that in such setups DM's relic abundance is determined via a duet of the \textit{superWIMP} mechanism \cite{Covi:1999ty,Feng:2003xh,Feng:2003uy} and freeze-in \cite{McDonald:1993ex,Hall:2009bx}, \textit{i.e.} the DM abundance today has a contribution from the direct freeze-in via the feeble portal interaction with SM in addition to its mediator's freeze-out and subsequent decay. 
This duet gives rise to an interesting viable parameter space and intriguing signals at colliders. It has been shown \cite{Garny:2018ali,Asadi:2023csb} that, thanks to the superWIMP mechanism responsible for DM abundance in the freeze-in regime, the viable DM mass range is bounded from above and is within reach of current and future collider searches. 
Furthermore, in contrast to the more studied freeze-out fermion-portal DM, the freeze-in regime models allow for both prompt and long-lived particle (LLP) collider signals.

A previous work \cite{Asadi:2023csb} focused on the signal of the right-handed (RH) charged lepton-portal at a muon collider. In this work, we extend that study to the remaining fermion portals. 
More explicitly, we will study the signals of models where DM couples to 1) a left-handed (LH) lepton $L$, 2) a LH quark $Q$, or 3) a RH up-type quark $u$ (the case of the RH down-type quark is similar to the RH up-type quark scenario). We note that different quantum numbers of the mediator give rise to qualitatively different signals at a muon collider, such as disappearing tracks for the LH lepton model or $R$-hadrons for the quark models.
In our analyses, we assume the DM interacts with all generations of SM fermions equally and leave studies of more structured interactions with different generations of SM particles to future work. 
Our goal is to delineate the reach of a high-energy muon collider in the parameter space of motivated DM models, while elaborating on details of the signal and the background that may prove useful in future development of a detector design.

We also augment our previous work by including the parton distribution function (PDF) of the muon component in the muon beam~\cite{Han:2020uid,Han:2021kes,AlAli:2021let,Garosi:2023bvq,Frixione:2023gmf,Marzocca:2024fqb}.
Unlike the effective vector approximation (EVA) for the PDF of gauge bosons, the muon PDF 
is not implemented 
in \texttt{MadGraph}~\cite{Ruiz:2021tdt}. 
We develop a methodology for performing this integration, which provides more accurate predictions for the signatures of our models.  
We find that the inclusion of this PDF is necessary for an accurate calculation of event distributions and will have non-trivial effects on our predictions for the reach of a muon collider in the parameter space of fermion-portal DM models.

The rest of this work is organized as follows. 
In \cref{sec:models} we introduce the different fermion-portal models. We also discuss their relic abundance calculations in the freeze-in limit and show that the phenomenologically viable DM masses are bounded from above.
We then discuss the implementation of the muon component PDF in the event generation pipeline in \cref{sec:PDF} and show that, while the total cross section does not change, the non-trivial muon component PDF affects the event distributions more significantly, which has important implications for a search for the fermion-portal models.
In \cref{sec:results} we study signals of different fermion-portal models at a muon collider. 
We propose a simple cut-and-count analysis in the prompt mediator decay part of the parameter space, which allows us to discover each model for the most of the kinematically--accessible mass range; we also report the event count rate in the long-lived mediator region. 
We conclude and discuss some implications of our results for a future detector design in \cref{sec:conclusion}.
In \cref{appx:masssplitting}, we discuss the small, yet crucial, mass splitting between mediators of different charges in models where the mediator is charged under SU(2)$_L$.

\section{Fermion-Portal Models in the Freeze-in Regime}\label{sec:models}

As one of the only a handful of renormalizable portals to a secluded dark sector, fermion-portal models constitute a very well-motivated target for various studies.
These models can give rise to a wide range of interesting phenomenology, including signals at various experimental searches (collider, direct and indirect detection, precision flavor experiments, etc.) \cite{Kile:2011mn,Agrawal:2011ze,Chang:2013oia,An:2013xka,DiFranzo:2013vra,Chang:2014tea,Garny:2014waa,Ibarra:2014qma,Kilic:2015vka,Agrawal:2015tfa,Bishara:2015mha,Blanke:2017fum,Chao:2017emq,Darme:2020ral,Maity:2023rez,Das:2024ghw}, novel ways of guaranteeing DM stability \cite{Batell:2011tc,Batell:2013zwa,Agrawal:2014aoa,Chen:2015jkt}, 
their potential role in baryogenesis and explaining the cosmological coincidence problem \cite{Hamze:2014wca,Agrawal:2016uwf,Dessert:2018khu}, and as part of the explanation of existing experimental anomalies \cite{Kile:2014jea,Kawamura:2020qxo,Bai:2021bau,Kawamura:2022uft,Acaroglu:2023cza}.

In this work, we focus on the regime in which the DM is produced via the freeze-in mechanism~\cite{McDonald:1993ex,Hall:2009bx}. 
In this regime, the DM has a very feeble coupling to the mediator and the SM, which can lead to interesting LLP signals at colliders.
The exact phenomenology, as well as the relevant areas of parameter space, of such models at muon colliders depends on the precise nature of the fermion portal. Ref.~\cite{Asadi:2023csb} studied the RH charged lepton-portal (``$e$ model"), carrying out studies of its prompt and long-lived particle signatures at a future muon collider. 
In this work, we expand on Ref.~\cite{Asadi:2023csb} to include the three other combinations: the LH lepton-portal (``$L$ model"), the RH quark-portal (``$u$ model"), and the LH quark-portal (``$Q$ model"). 
Specifically, we introduce the details of each model and calculate the values of the mediator-DM coupling that give the correct abundance as a function of mediator and DM masses. Determining this coupling, in turn, fixes the lifetime of the mediators in each model, which drives the model's phenomenology at a future muon collider.

The basic cosmological history behind all four models is the same. In all the scenarios, we will assume that the reheat temperature of the universe is above the mediator mass and the gauge interactions of the mediator keep it in equilibrium with the SM bath. As the universe cools and expands, the mediator goes through the standard freeze-out process, and subsequently decays to the DM $\chi$ and the $\phi$ mediator's SM counterpart. Meanwhile, the DM particle $\chi$ is directly populated through freeze-in via the feeble coupling $\lambda \ll 1$, where the contribution is proportional to the decay rate of the mediator $\phi$ into DM and SM~\cite{Asadi:2023csb}.

Through this cosmological history, we can gain some general intuition for the dependence of the relic abundance, and therefore the required $\lambda$, on $m_\chi$ and $m_\phi$.
In much of the parameter space, the model's relic abundance today is dominated by the direct freeze-in of DM. 
For a fixed $m_\phi$, the required $\lambda$ decreases as $m_\chi$ increases. This can be understood as follows. Fixing $m_\phi$ fixes the number of $\chi$ particles that result from the freeze-out and decay of the mediator $\phi$. Since the energy density of the DM $\chi$ increases with increasing $m_\chi$, the value of $\lambda$, which controls the freeze-in contribution to the DM abundance, must decrease to compensate. 
For a fixed $m_\chi$, the required $\lambda$ increases as $m_\phi$ increases. An increase in $m_\phi$ leads to a decrease in the asymptotic contribution of direct freeze-in to the final number density (see Ref.~\cite{Hall:2009bx} or the appendix of Ref.~\cite{Asadi:2023csb}). This decrease is compensated by an increase in $\lambda$ in order to populate the observed relic abundance.

As we go to larger values of $m_\phi$ with a fixed $m_\chi$, eventually the freeze-out and decay of $\phi$ will result in an excessive abundance of $\chi$ particles that overcloses the universe, even neglecting the direct freeze-in contribution. 
As a result, for each $m_\chi$ there is an upper bound on the value of $m_\phi$ that allows us to explain the observed DM abundance today~\cite{Planck:2018vyg} without overclosing the universe. 
Our entire relic abundance calculation is done with the assumption of the eventual $\phi\rightarrow \chi+{\rm SM}$ decay, which requires $m_\phi \gtrsim m_\chi$; this limits the viable mediator mass range from below.
In addition, there are lower limits on $m_\chi$ which arise from structure formation and require $m_\chi$ to be larger than ${\cal O}(10{~\rm keV})$~\cite{DEramo:2020gpr},\footnote{These bounds can be relaxed if the freeze-in DM is a sub-dominant component of the total DM abundance \cite{DEramo:2020gpr}.} as well as collider bounds on the mediator \cite{ATLAS:2017tny, ATLAS:2019lff,CMS:2019zmd,CMS:2019ybf,ATLAS:2020syg,CMS:2020bfa,ATLAS:2020wjh, ATLAS:2022pib, ATLAS:2023tbg, CMS:2024qys}. Therefore, the viable $m_\phi$ vs. $m_\chi$ parameter space is bounded from all directions.

We begin with a brief review of the results from solving the Boltzmann equations for the $e$ model before calculating all the relevant quantities for the remaining fermion-portal models: $L$, $u$, and $Q$ models. 
The list of Feynman diagram topologies relevant for these calculations can be found in \cref{tab:diagrams}. 
The 2-to-2 processes are relevant for the annihilation and freeze-out of the mediator, while the 2-body mediator decay is the dominant source of DM freeze-in.

\begin{table}[h]
    \centering
    \resizebox{\columnwidth}{!}{
    \begin{tabular}{c||c|c|c|c}
     Topology & $e$ model & $L$ model & $u$ model & $Q$ model \\
    \hline
    \multirow{3}{*}{\begin{minipage}{0.15\textwidth}
      \includegraphics[height=17mm]{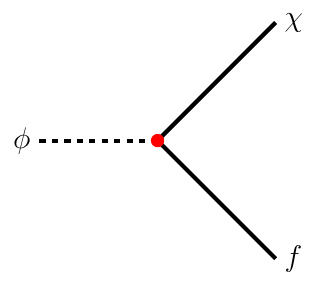}
    \end{minipage}} &$\phi_e\to e\chi$ & $\phi^0_L\to \nu \chi$ & $\phi_u\to u\chi$ & $\phi_Q^{2/3} \to u\chi$ \\
    & & $\phi_L^-\to \ell^-\chi$ & &$\phi^{-1/3}_Q\to d \chi $ \\ 
    & & & & \\
\hline
     \multirow{3}{*}{\begin{minipage}{.15\textwidth}
      \includegraphics[height=17mm]{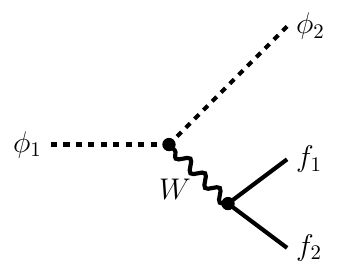}
    \end{minipage}} 
     & & & & \\
        & &$\phi_L^-\to\phi_L^0\ell^- \bar\nu$ & & $\phi_Q^{2/3}\to\phi_Q^{-1/3} e^+ \nu_e$\\   & & & & \\
\hline
\hline
     \multirow{3}{*}{\begin{minipage}{.15\textwidth}
      \includegraphics[height=17mm]{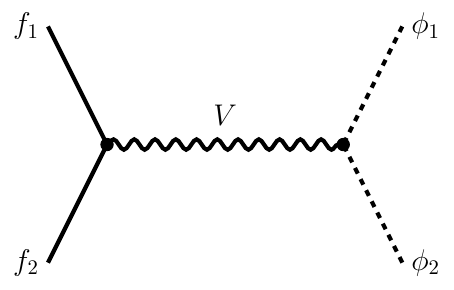}
    \end{minipage}} 
     & & $\mu\mu \to \phi_L^+ \phi_L^-$ & & $\mu\mu \to \phi_Q^{2/3} \phi_Q^{-2/3}$ \\
    & $\mu\mu \to \phi_e \phi_e$ & $\mu\mu \to \phi_L^0 \phi_L^0$ & $\mu\mu \to \phi_u \phi_u$ & $\mu\mu \to \phi_Q^{1/3} \phi_Q^{-1/3}$ \\  
    & & & & \\
\hline
     \multirow{3}{*}{\begin{minipage}{.15\textwidth}
      \includegraphics[height=17mm]{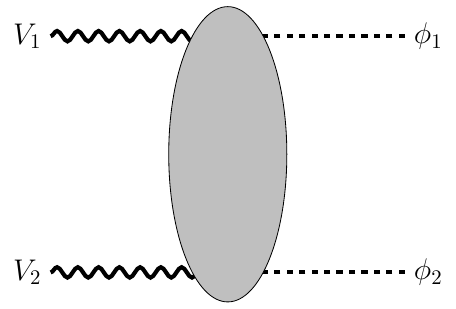}
    \end{minipage}} 
       & & $VV \to \phi_L^+ \phi_L^-$ & & $VV \to \phi_Q^{2/3} \phi_Q^{-2/3}$ \\
         & $VV \to \phi_e \phi_e$ & $VV \to \phi_L^0 \phi_L^0$ & $VV \to \phi_u \phi_u$ & $VV \to \phi_Q^{1/3} \phi_Q^{-1/3}$ \\  
         & & $VV \to \phi_L^0 \phi_L^\pm$ & & $VV \to \phi_Q^{2/3} \phi_Q^{1/3}$  
    \end{tabular}
    }
     \caption{Decay and production channels for the mediators in fermion-portal dark matter that are relevant for both the Boltzmann equations and collider studies. For each topology, we report the specific process in each model in different columns.  The {\bf \color{red}red} dot in the first row denotes the freeze-in coupling $\lambda$, while the {\bf black} dots denote the SM gauge couplings. The first two rows denote possible decay topologies, while the last 2 rows give the annihilation topologies that are relevant for the Boltzmann equations, as well as for the collider production of the mediators. Here $V$ stands for any of SM's electroweak gauge bosons; the subscript on each mediator denotes the model to which they correspond.  }
    \label{tab:diagrams}
   \end{table}

\subsection{Right-Handed Leptons Portal ($e$ model)}\label{subsec:RHlep}

For completeness, we briefly discuss the RH charged lepton model studied in Ref.~\cite{Asadi:2023csb} and summarize the main features. 
In this scenario, we add to the SM a scalar mediator $\phi_e$, which shares the same SM gauge charges as the RH charged leptons, and a SM neutral fermion $\chi$,
\begin{equation}
\mathcal{L}_{\rm int}^{e \textrm{ model}} \supset - m_\chi \bar{\chi}_\alpha \chi^\alpha - m_\phi^2 |\phi_e|^2 - \lambda_{i,\alpha} \phi_e \bar{e}_i \bar{\chi}_{\alpha}\, ,
    \label{eq:L}
\end{equation}
where $\alpha ~ (i)$ is the DM (SM) fermion flavor index and $\bar e$ is the RH charged SM lepton. $m_\phi$ and $m_\chi$ denote the mediator and DM masses, respectively. $\lambda$ is the DM Yukawa coupling that determines its freeze-in relic abundance; for simplicity, we assume the DM has the same coupling to all generations of SM leptons. 
The choice of the flavor texture can affect the relic abundance calculation and collider signatures. 

\begin{figure}
    \centering
    \includegraphics[width=0.49\linewidth]{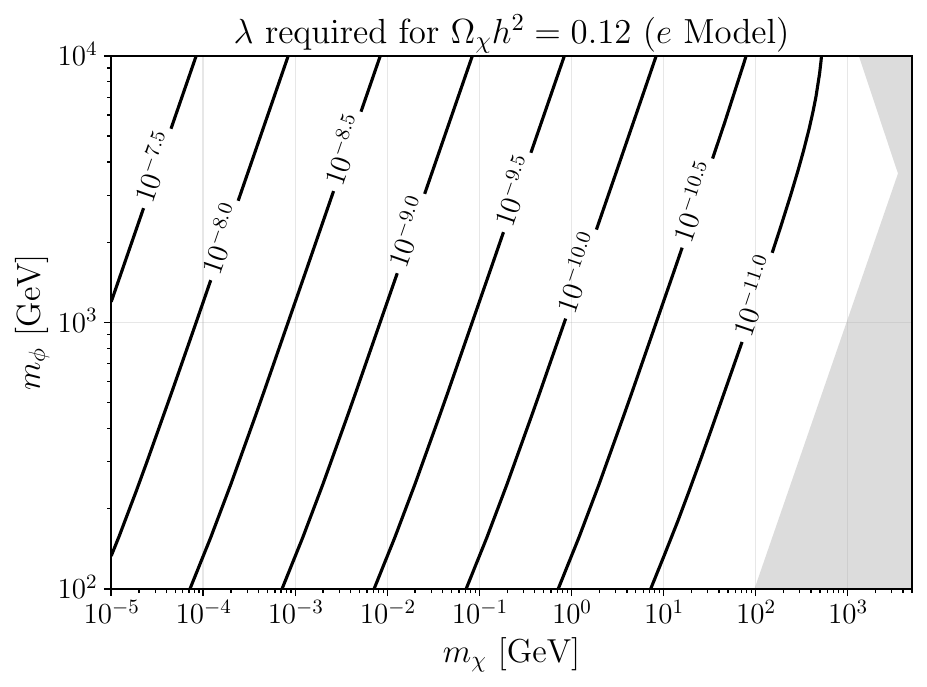}
    \includegraphics[width=0.49\linewidth]{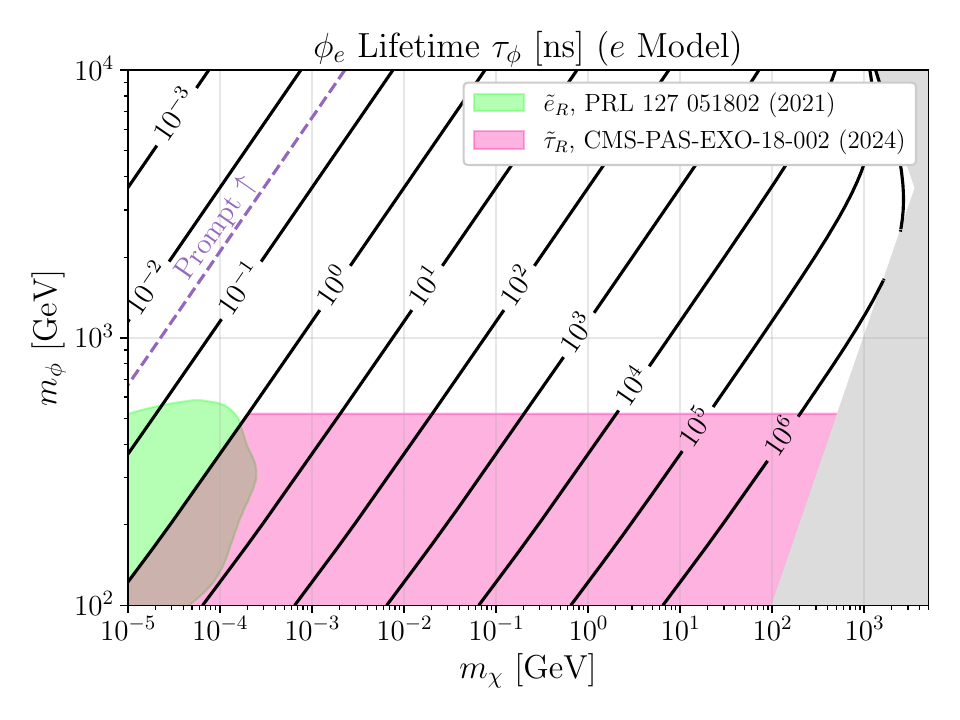}
    \caption{\underline {$e$ model:} ({\bf left}) The DM Yukawa coupling, $\lambda$, that gives rise to the correct relic abundance today. Here, we have taken the coupling $\lambda_i\equiv\lambda$ to be the same for all fermion flavors. 
    The interplay of the freeze-in and superWIMP mechanisms gives rise to an upper bound on DM mass.
    The top right gray region is excluded from overclosure constraints, while the bottom right gray region is kinematically not allowed in our scenario. Strong astrophysical bounds rule out $m_\chi$ below $\sim$10 keV~\cite{DEramo:2020gpr}, giving rise to a completely-bounded parameter space. ({\bf right}) $\phi_e$ lifetime as a function of $m_\phi$ and $m_\chi$. The {\bf\textcolor{xkcdNeonGreen}{green}} and {\bf\textcolor{xkcdNeonPink}{pink}} regions indicate current LHC slepton constraints from Ref.~\cite{ATLAS:2020wjh} and Ref.~\cite{CMS:2024qys}, respectively. For all the points above (below) the dashed purple line, the mediator appears as a promptly-decaying (long-lived) particle in colliders. }
    \label{fig:lambda_parameter}
\end{figure}

The DM is also populated through the freeze-out production ($\phi_e^+\phi_e^-\leftrightarrow\cal{F}$, where ${\cal F}\in [\ell\bar\ell,\,\gamma\gamma,\,q\bar q,\,ZZ,\,WW]$) and subsequent decay of the mediator ($\phi_e^\pm\to\ell^\pm\chi$). Solving the Boltzmann equations for the DM relic abundance gives an upper bound on the DM mass of
\begin{equation}
m_\chi^{e \textrm{ model}} \lesssim 3.6~\mathrm{TeV}\, ,
    \label{eq:DMupper}
\end{equation}
see Ref.~\cite{Asadi:2023csb} for a complete treatment of the Boltzmann equations. The Yukawa coupling that gives rise to the correct relic abundance today and the mediator lifetime, as a function of mediator and DM mass, are shown in Fig.~\ref{fig:lambda_parameter}.

\subsection{Left-Handed Leptons Portal ($L$ model)}\label{subsec:LHlep}

The next portal we consider is the LH leptons portal, which has the following Lagrangian:
\begin{equation}
    {\cal L}_{\rm int}^{L \textrm{ model}}\supset -\lambda_{i}\Phi_L^\dagger L_i \bar\chi  - m_\chi \bar{\chi} \chi - m_\phi^2 \Phi^\dagger_L \Phi_L + h.c.\, .
    \label{eq:L_L}
\end{equation}
Here, $L$ is a left-handed SM lepton, $i$ denotes the SM lepton's flavor indices, $\Phi_L = \left( \begin{matrix} \phi_L^0 \\ \phi_L^- \end{matrix} \right)$ is a complex scalar doublet with the same charges as $L$ under the SM gauge group, $\chi$ and $\bar{\chi}$ are the LH and RH DM particles, and the subscript $L$ on the new fields refer to the fact that they belong to the $L$ model. 
In contrast to the $e$ model, we now have two mediators, a charged and neutral mediator, which are nearly degenerate in mass, split only by gauge interactions at one-loop~\cite{Cirelli:2005uq} (see Appendix~\ref{appx:masssplitting} for more detail). The existence of a charged mediator introduces additional channels to the early-Universe production of the DM as there are new Feynman diagrams involving the electroweak couplings.  

\begin{figure}
            \centering
            \includegraphics[width=0.6\linewidth]{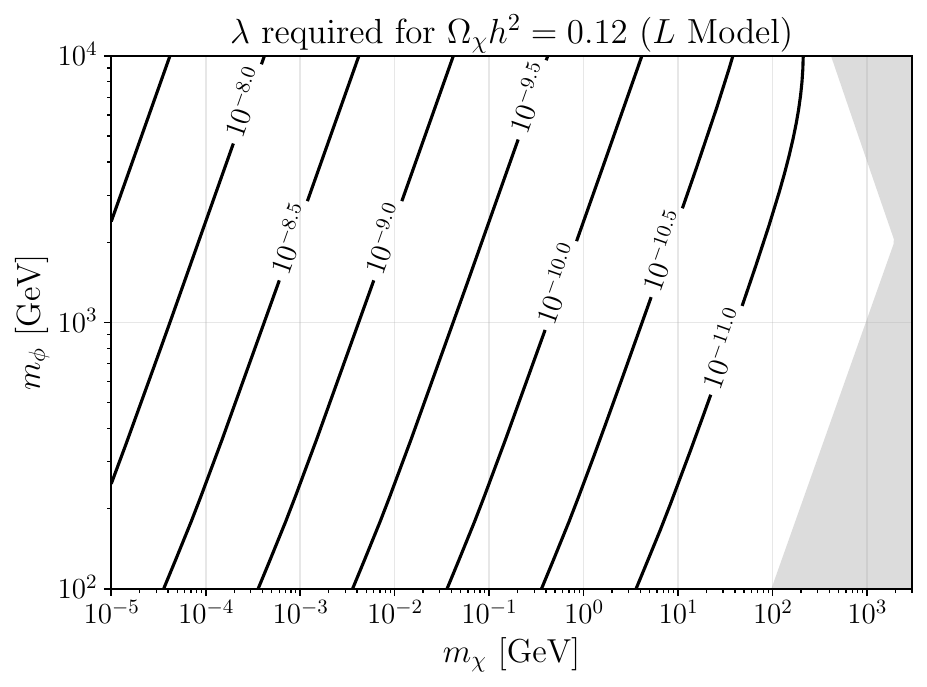}
            \caption{\underline{$L$ model:} The DM Yukawa coupling, $\lambda$, that gives rise to the correct relic abundance today. Here, we have taken the coupling $\lambda_i\equiv\lambda$ to be the same for all fermion flavors. The top right gray region is excluded from overclosure constraints, while the bottom right gray region is kinematically not allowed in our scenario. Strong astrophysical bounds rule out $m_\chi$ below about $\sim$10 keV~\cite{DEramo:2020gpr}, giving rise to a completely-bounded parameter space.}
            \label{fig:lambda_L}
\end{figure}

As before, we will assume the same Yukawa coupling $\lambda_i=\lambda$ between the new particles and all generations of SM's $L$ fermions.
We solve the Boltzmann equations for this model to get the relic abundance today as a function of the coupling $\lambda$. 
Figure~\ref{fig:lambda_L} shows the required values of the $\lambda$ coupling that give rise to the observed relic abundance today.
Note that compared to the $e$ model, the values of $\lambda$ are slightly smaller. This can be understood as a consequence of the additional DM production channels arising from its non-trivial SU(2)$_L$ transformation. 

\begin{figure}
    \centering
    \resizebox{\columnwidth}{!}{
    \includegraphics[width=0.8\linewidth]{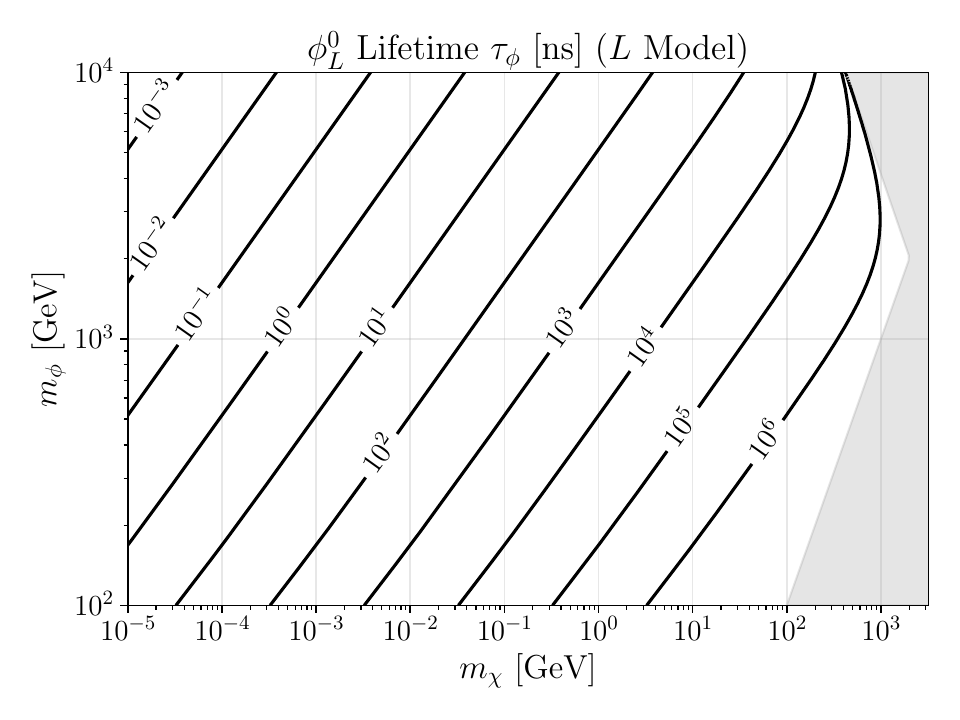}
    \includegraphics[width=0.8\linewidth]{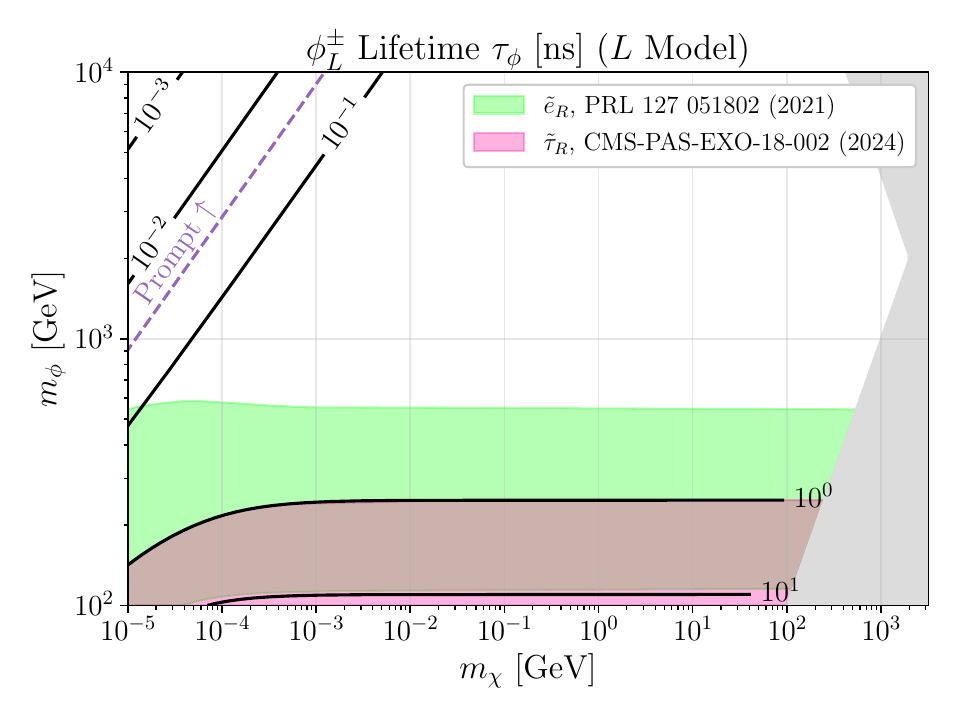}
    }
    \caption{\underline{$L$ model:} The lifetime of the neutral (\textbf{left}) and the charged (\textbf{right}) mediator. For every point in the parameter space we assume the freeze-in coupling $\lambda$ is set to the value that gives rise to the observed DM abundance today, see~\cref{fig:lambda_L}. Both mediators span a large range of lifetimes, from promptly-decaying at a collider (above the dashed purple line), to behaving as an LLP across the rest of the parameter space. 
    The {\bf\textcolor{xkcdNeonGreen}{green}} and {\bf\textcolor{xkcdNeonPink}{pink}} regions indicate current LHC slepton constraints from Ref.~\cite{ATLAS:2020wjh} and Ref.~\cite{CMS:2024qys}, respectively.  The neutral mediator can decay only via the freeze-in coupling $\lambda$, while the charged mediator can also decay to the neutral one via an off-shell $W$, see the text for further details about the shape of contours in these plots. }
    \label{fig:lifetime_L}
\end{figure}

The combination of the two excluded regions discussed above gives rise to an upper bound on DM mass:
\begin{equation}
    m_\chi^{L \textrm{ model}} \lesssim 2.0~\mathrm{TeV}.
    \label{eq:boundmL}
\end{equation}
Using the values of $\lambda$ in \cref{fig:lambda_L}, we can calculate the lifetime of each mediator for each point in the parameter space.
The result of this calculation is shown in~\cref{fig:lifetime_L}. 
For lifetimes less than $\tau_\phi\sim 10^{-1.5}$ ns, the mediator is short-lived enough to appear as a promptly-decaying particle at a collider, while for lifetimes $\tau_\phi \gtrsim 10^{-1.5}$ ns, we will have LLP signatures. 

The difference in shapes of the lifetime contours between the neutral and charged mediator is due to the available decay channels. The neutral mediator in this setup can only decay via the small freeze-in coupling ($\phi_L^0\to\nu\chi)$, see the first row diagram in \cref{tab:diagrams}. 
The charged mediator $\phi_L^-$, on the other hand, has two decay channels which leads to qualitatively different features. The first channel is the analogous $\phi_L^\pm\to\ell^\pm\chi$ decay via the small freeze-in coupling. The second arises via an off-shell $W$ boson ($\phi_L^\pm\to \phi_L^0\ell^\pm\bar\nu$), see the second row diagram in \cref{tab:diagrams}.

\begin{figure}
            \centering
            \includegraphics[width=0.99\linewidth]{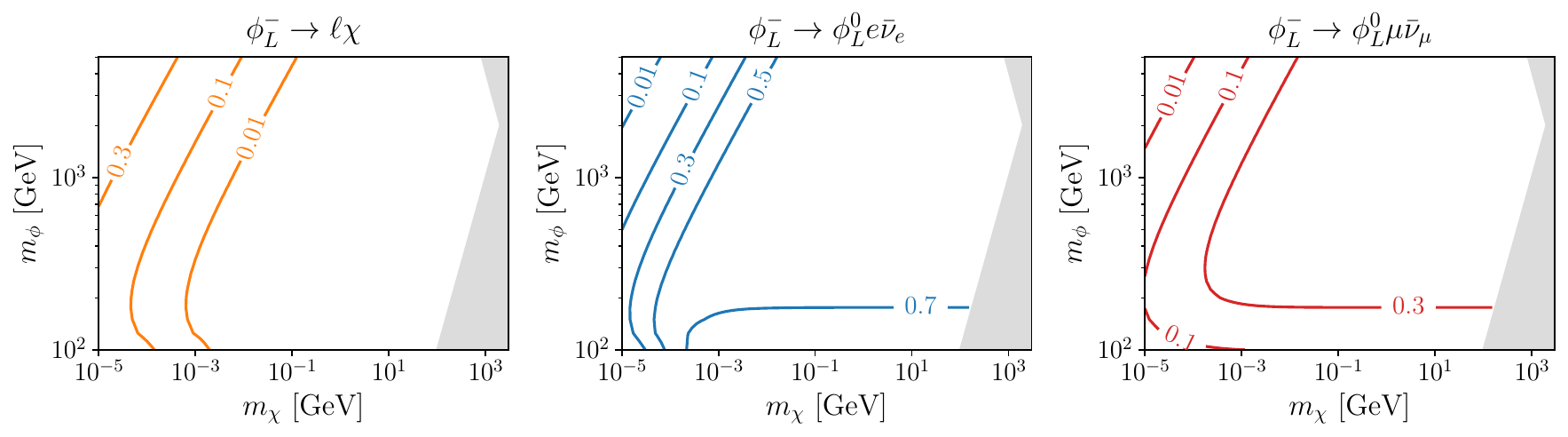}
            \caption{\underline{$L$ model:} Branching ratio of the charged mediator $\phi_L^-$ to different channels: DM and a charged lepton (\textbf{left}), neutral mediator $\phi_L^0$ accompanied by a neutrino and a soft electron (\textbf{middle}) or muon (\textbf{right}). Note the difference between the latter two channels arises from the phase space suppression since the mass splitting between the two mediators is comparable to the muon mass. We find that in the prompt region the mediator predominantly decays to DM and a charged lepton, while in the LLP region it decays predominantly to the neutral mediator and soft leptons, giving rise to a disappearing track signal as elaborated in Section~\ref{subsubsec:llp_L}.}
            \label{fig:BR_L}
\end{figure}

The phenomenology of this model at colliders is affected by the branching ratio of the charged mediator into different possible final states (see the first two rows from \cref{tab:diagrams}).\footnote{There is also a 4-body decay channel that is at most at the percent level in the parameter space of the model, and can be safely neglected in our study.}
The branching ratios for different possible decays are shown in \cref{fig:BR_L}. 
In the prompt region of the parameter space, \textit{i.e.} $\tau_\phi\lesssim 10^{-1.5}$ ns, the charged mediator overwhelmingly decays to a DM and a charged lepton. Thus, the main signal of the model at a collider in this region is a pair of charged leptons (as the mediator is pair produced) and missing energy.
In the LLP regime, the mediator mostly decays to the neutral mediator and soft undetectable leptons.
When this decay channel becomes dominant, the lifetime of the mediator becomes independent of the DM mass or its $\lambda$ coupling, as shown in~\cref{fig:lifetime_L}. 

\subsection{Right-Handed Quarks Portal ($u$ model)}\label{subsec:RHq}

The next class of models we consider are the RH up-type quark-portal\footnote{The RH down-type quark-portal has qualitatively similar signals and are not studied separately in this work.}
\begin{equation}
    {\cal L}_{\rm int}^{u \textrm{ model}}\supset 
   -\lambda_{i}\phi_u\bar u_i \bar\chi - m_\chi \bar{\chi} \chi  - m_\phi \phi^\dagger_u \phi_u + h.c.,
   \label{eq:L_u}
\end{equation}
where now the mediator $\phi_u$ is an $\textrm{SU}(3)_c$ fundamental, charge 2/3 scalar. Here the subscript $u$ refers to the fact that this is the $u$ model mediator. 
As the mediator now carries color, this model has a distinct phenomenology from the lepton models.
For example, the freeze-out of the mediator receives contributions from color-mediated annihilation channels, which have larger cross sections than the previously considered setups. 
As a result, the mediator will have a higher annihilation rate, which results in a smaller asymptotic freeze-out abundance for DM after the mediator decays. 
To compensate this, the required $\lambda$ couplings to get the right relic abundance today, and the upper bound on DM mass, are larger than the lepton-portal models.

\begin{figure}
    \centering
    \resizebox{\columnwidth}{!}{
    \includegraphics[width=0.6\linewidth]{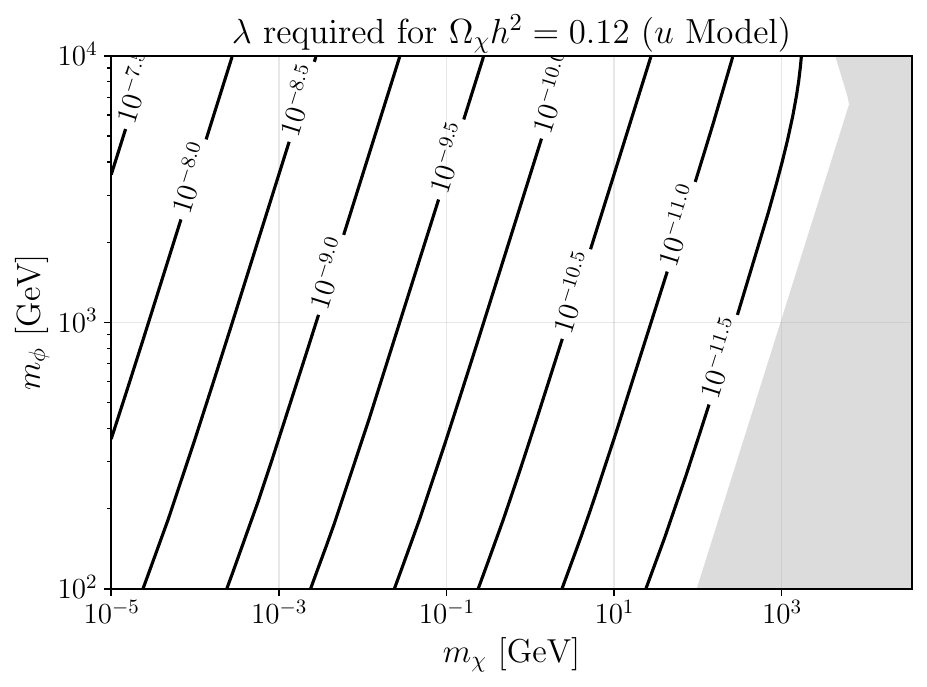}
    \includegraphics[width=0.6\linewidth]{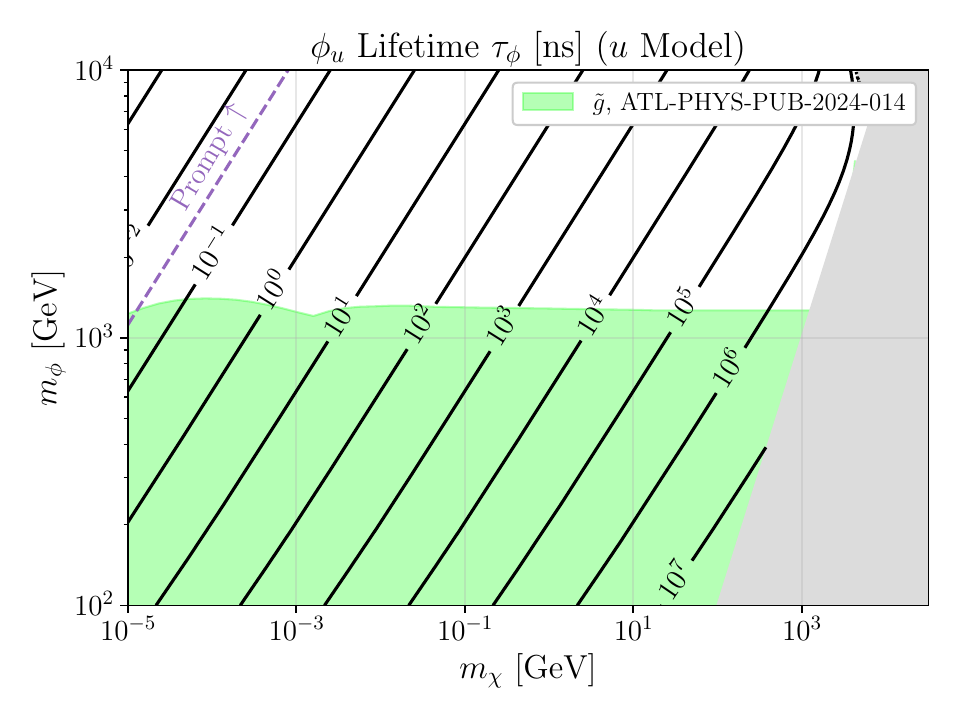}
    }
    \caption{
    \underline{$u$ model:} \textbf{(left)} The required freeze-in coupling $\lambda$ from Eq.~\eqref{eq:L_u} of the RH quark-portal model to explain the observed DM relic abundance from a combination of the freeze-in and the superWIMP mechanisms. The interplay of these mechanisms gives rise to an upper bound of on DM mass. \textbf{(right)} The lifetime of the color-charged mediator in the RH quark-portal model using the values of $\lambda$ that yields the correct relic abundance today. The mediator's lifetime spans a large range, from promptly-decaying at a collider (top-left part of each plot), to behaving as an LLP across the rest of the parameter space. The only available decay channel of the mediator is via the freeze-in coupling $\lambda$ to SM quarks and DM. Current bounds from LLP searches at ATLAS are shown in {\bf\textcolor{xkcdNeonGreen}{green}} \cite{ATL-PHYS-PUB-2024-014}. }
    \label{fig:lambda_lifetime_u}
\end{figure}

The required values of $\lambda$ coupling that gives rise to the observed DM relic abundance today are shown in~\cref{fig:lambda_lifetime_u}, from which we can read off the upper bound on the DM mass
\begin{equation}
    m_\chi^{u \textrm{ model}} \lesssim 6.7{~\rm TeV}.
    \label{eq:boundmu}
\end{equation}
It should be noted that given the current bounds on the color-charged mediator $\phi_u$, the freeze-out happens in the deconfined phase of QCD and at high enough temperatures that we can treat the strong interactions as perturbative.

Finally, using the values of $\lambda$ from~\cref{fig:lambda_lifetime_u} we can calculate the lifetime of the mediator in this setup. 
The result is shown in the same figure. 
We again find the mediator can either decay promptly or be an LLP in different parts of the parameter space, each requiring different search strategies, as will be discussed in Section~\ref{subsec:collideru}.

The color-charged nature of the mediators in the $u$ model (and the following $Q$ model) also gives rise to distinct collider signatures compared to the lepton models. In the LLP part of the parameter space, the main signal of the model will be the so-called $R$-hadrons \cite{Farrar:1978xj}. 
These are jets made from a heavy new color-charged LLP, the mediator $\phi_u$ in this case, which forms color-neutral objects with SM fermions pulled out of the vacuum, and can have distinctive features in colliders. We will discuss these features in more detail in Sec.~\ref{subsec:collideru}.

\subsection{Left-Handed Quarks Portal ($Q$ model)}\label{subsec:LHq}

The final class of models we consider in this work is the LH quarks portal, the $Q$ model, which has the following Lagrangian
\begin{equation}
    {\cal L}_{\rm int}^{Q \textrm{ model}}\supset -\lambda_{i}\Phi_Q^\dagger Q_i \bar\chi - m_\chi \bar{\chi} \chi - m_\phi \Phi^\dagger_Q \Phi_Q + h.c. ,
    \label{eq:L_Q}
\end{equation}
where now the mediator $\Phi_Q = \left( \begin{matrix}  \phi_Q^{2/3} \\ \phi_Q^{-1/3} \end{matrix} \right)$ has the same gauge charges as the LH $Q$ fermions in SM, indicated by the subscript $Q$ on the new fields. 
Again, we assume the mediator couples democratically to all SM generations.
This model contains a combination of the features seen in the $u$ model and the $L$ model. As in the $u$ model, the mediators are now color-charged and as in the $L$ model, there are two nearly mass-degenerate mediators. The primary difference is that there are now {\it two charged} mediators, as opposed to one charged and one neutral.

We calculate the mass splitting between the two charged mediators using Ref.~\cite{Cirelli:2005uq} and find that the charge $1/3$ mediator is the lighter of the two by about 0.1 GeV (see Appendix~\ref{appx:masssplitting}). 
Thus, the lighter mediator can only decay as $\phi_Q^{-1/3}\to \chi  d$ via the Yukawa interaction in Eq.~\eqref{eq:L_Q}, while the heavier mediator has two decay channels: $\phi_Q^{2/3}\to\chi u$ and an additional decay channel to the lighter mediator via an off-shell $W$, $\phi_Q^{2/3}\to \phi_Q^{-1/3} e \nu$. 

\begin{figure}
            \centering
            \includegraphics[width=0.6\linewidth]{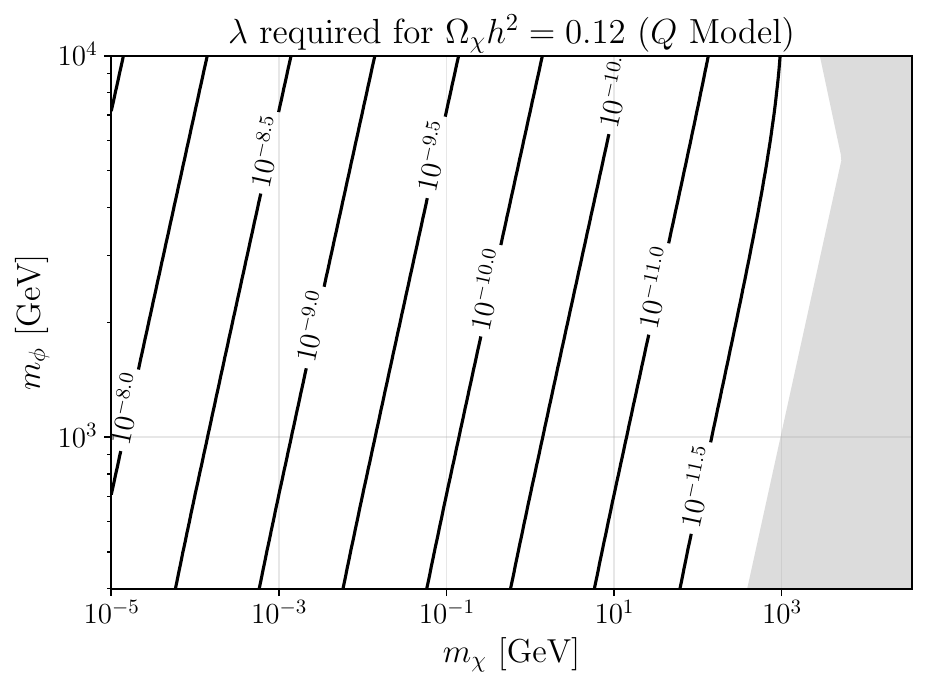}
            \caption{\underline{$Q$ model:} The required freeze-in coupling $\lambda$ from Eq.~\eqref{eq:L_Q} of the LH quark-portal model to explain the observed DM relic abundance from a combination of the freeze-in and the superWIMP mechanisms. The interplay of these mechanisms gives rise to an upper bound on DM mass.}
            \label{fig:lambda_Q}
\end{figure}

\begin{figure}
            \centering
            \resizebox{\columnwidth}{!}{
            \includegraphics[width=0.6\linewidth]{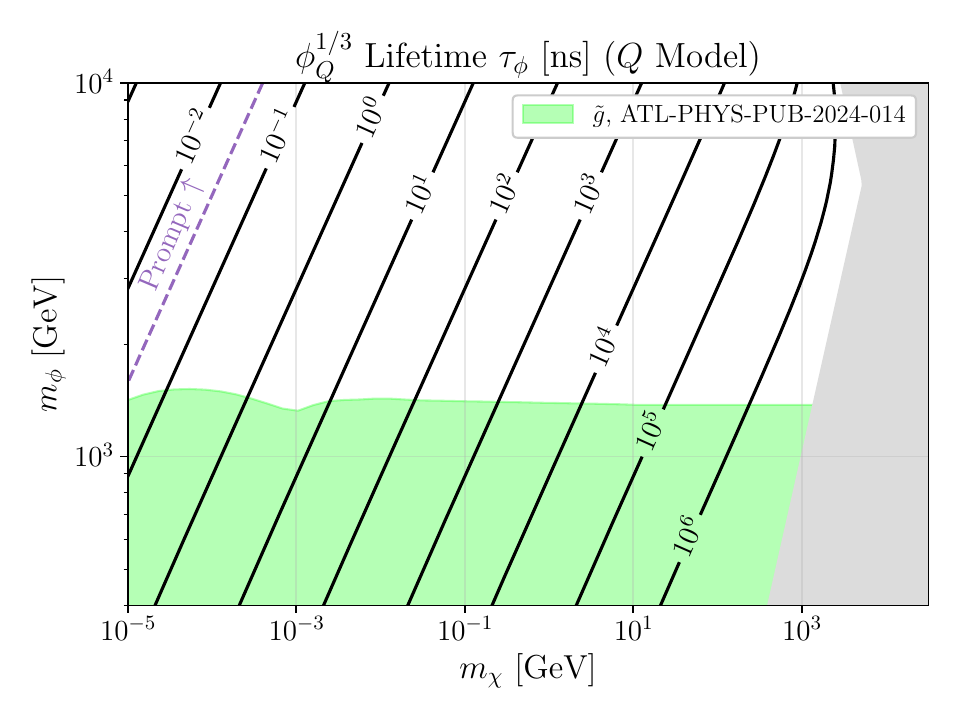}
            \includegraphics[width=0.6\linewidth]{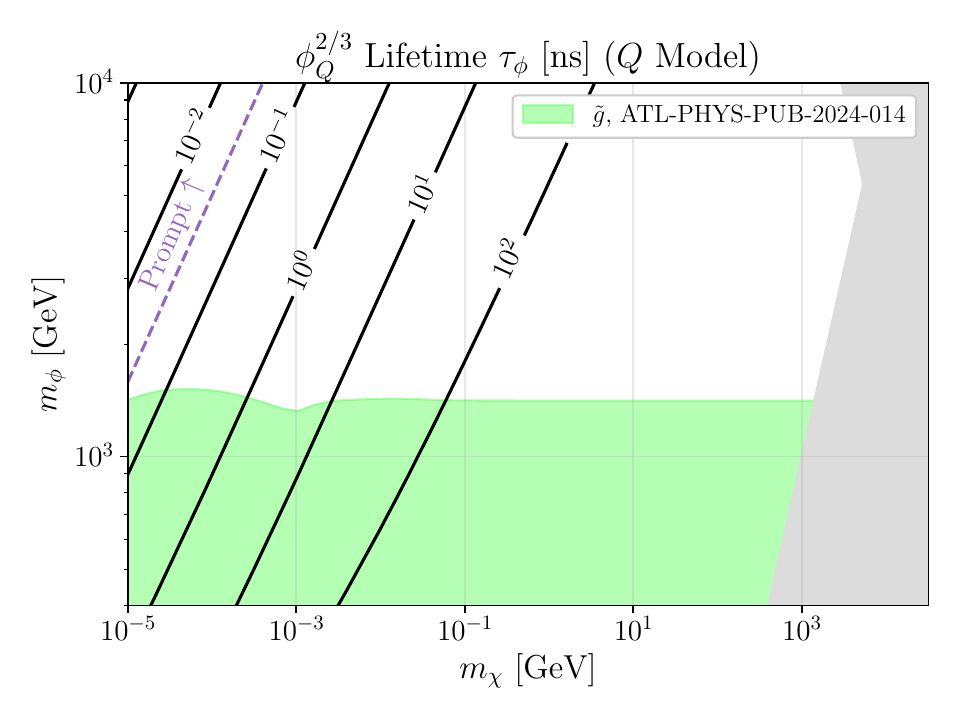}
            }
            \caption{
            \underline{$Q$ model:}
            The lifetime of the lighter (\textbf{left}) and the heavier  (\textbf{right}) charged mediators in the LH quark-portal model; for every point in the parameter space we assume the freeze-in coupling $\lambda$ is set to the value that gives rise to the observed DM abundance today, see~\cref{fig:lambda_Q}. Both mediators span a large range of lifetimes, from promptly-decaying at a collider (top-left part of each plot), to behaving as an LLP across the rest of the parameter space. The lighter mediator can decay only via the freeze-in coupling $\lambda$, while the heavier one can also decay to the lighter one via an off-shell $W$, see the text for further details. Current bounds from LLP searches at ATLAS are shown in {\bf\textcolor{xkcdNeonGreen}{green}} \cite{ATL-PHYS-PUB-2024-014}.}
            \label{fig:lifetime_Q}
\end{figure}

We can again use the relic abundance calculation and find the values of $\lambda$ that yield the correct DM energy density today for every point on the parameter space and, subsequently, use these values of $\lambda$ to calculate the lifetime of each mediator.
Contours of constant $\lambda$ (mediator lifetime) on the parameter space of this model are shown in~\cref{fig:lambda_Q} (\cref{fig:lifetime_Q}) and show analogous behavior to the $L$ model. 
These figures show that the upper bound on the DM mass in this setup is
\begin{equation}
    m_\chi^{Q \textrm{ model}} \lesssim 5.4~\mathrm{TeV}.
    \label{eq:boundmQ}
\end{equation}

\begin{figure}
            \centering
            \includegraphics[width=0.7\linewidth]{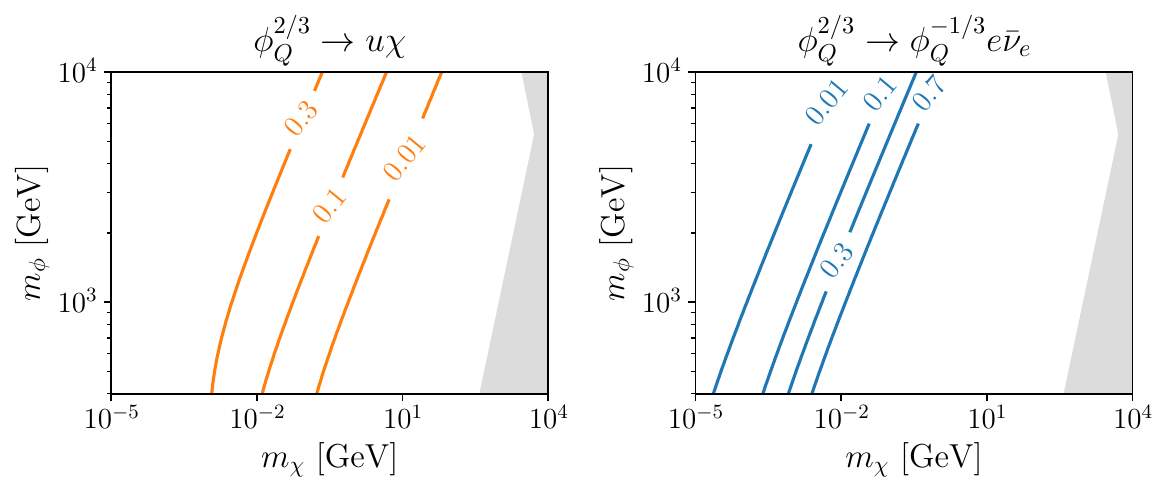}
            \caption{\underline{$Q$ model:} Branching ratio of the heavier, charge-2/3 mediator $\phi_Q^{2/3}$ to different channels: DM and a light up-type quark (\textbf{left}), or a lighter charge-1/3 mediator accompanied by a neutrino and a soft electron (\textbf{right}). The analogous 3-body decay with a soft muon is vanishingly small due to the smaller mass splitting between $\phi_Q^{2/3}$ and $\phi_Q^{1/3}$ than in the $L$ model.
            We find that in the prompt region the mediator predominantly decays to DM and a light quark, while in the LLP region it decays predominantly to the neutral mediator and a soft electron, giving rise to a disappearing track signal as elaborated in Section~\ref{subsubsec:llp_Q}. }
            \label{fig:BR_Q}
\end{figure}

To reiterate, the $Q$ model setup has the main qualitative features of the previous models discussed: the mediators are both color-charged and nearly mass-degenerate. 
As a result, the collider signatures will be a combination of the signatures of the previous models. 
For the heavier mediator, these signatures are determined by its branching ratio to different channels, as shown in \cref{fig:BR_Q}.
In the prompt region we have two charged color mediators that decay to jets and missing energy. 
Similar to the $L$ model, in the LLP region the heavy mediator dominantly decays to the lighter (color-charged and charged) mediator and soft SM lepton (via an off-shell $W$), and the light mediator in turn subsequently decays to a jet and missing energy. In this region, the model gives rise to $R$-hadron signals, as discussed in the $u$ model. A more detailed study of the collider signatures is presented in Sec.~\ref{subsec:colliderQ}.

This completes the review of the models in the current study. 
We found that each model includes new mediators, at least a part of whose viable mass ranges are within the reach of future colliders.
In the rest of this work we plan on studying signals of these models at a future 10~TeV muon collider.
Before that, however, we will discuss implementation of the muon component PDF in \texttt{MadGraph}, which affects the collider studies of these models.

\section{Muon Colliders and Muon PDFs}\label{sec:PDF}

We now turn to the production of fermion-portal mediators at a high-energy muon collider.
As mentioned above, we will focus our attention on a $10~\textrm{TeV}$ center-of-momentum energy collider, assuming $10~\textrm{ab}^{-1}$ total integrated luminosity. 
This is approximately the energy of the ``site-filling'' design envisioned at Fermilab~\cite{Black:2022cth}, and $10~\textrm{TeV}$ is frequently used in phenomenological studies, especially in the context of new physics, so it is a useful point for comparison. 

At multi-TeV energies, the logarithmic corrections associated with soft and collinear divergences become more significant. It is thus desirable to re-sum these large logarithms to improve the perturbative expansion. This brings about the PDF formalism for high-energy lepton beams, in complete analogy to PDFs in QCD with the key difference that the boundary conditions for the DGLAP evolution can be computed perturbatively, making the entire problem a computational one.
At the high energies under consideration, the entire SM field content and interactions must be taken into account, including the polarization effects resulting from the parity breaking in the weak interactions~
\cite{Chen:2016wkt, Bauer:2018arx, Han:2020uid, Han:2021kes, AlAli:2021let, Garosi:2023bvq, Frixione:2023gmf, Marzocca:2024fqb}.

The results in the following sections make use of the muon PDFs calculated in Ref.~\cite{Garosi:2023bvq}. These ``LePDF'' results are published in a format that closely follows the format of the proton PDFs in the LHAPDF package~\cite{Buckley:2014ana}, with minimal changes to account for the polarization information in the muon PDFs. This is done in anticipation of the eventual need to interface the muon PDFs directly with event generators for precision calculations of high-energy lepton collisions.

Unfortunately, there are several hurdles to a direct implementation of the muon PDFs in event generator codes such as \texttt{MadGraph}.
First, due to the explicit parity breaking in the weak interactions, the PDFs must be split into different chiralities, in contrast to proton PDFs.
The muon components of the PDFs also have an integrable singularity at $x \to 1$, which cannot be evaluated numerically.
Finally, there is the presence of novel ``interference-PDFs'', which cannot be read by existing generators. 

Of these, the last is the most challenging to manage with existing codes.
Mixed PDFs allowing for mixtures of transversely polarized $Z$ bosons and photons and longitudinal $Z$ bosons and the Higgs must be included to account for the fact that processes with these initial states can interfere~\cite{Ciafaloni:2000gm, Ciafaloni:2005fm, Chen:2016wkt}.
From the perspective of the PDF, the interference PDFs are crucial to include because the DGLAP equations do not respect the boundary conditions with no mixing.
The necessary mixed states are included in the DGLAP equations used to solve for the muon PDFs (this leads to one of the biggest differences in comparison to the effective vector approximation), and the resulting interference PDF is included in the grids exported by LePDF~\cite{Garosi:2023bvq}. 
The challenge is that these mixed PDFs must be convolved with the {\em interference} term in the squared matrix element. At the moment, the necessary formalism involving mixed states is not implemented in event generators such as \texttt{MadGraph5}, and can only be done analytically.\footnote{
One potential workaround is to convolve the mixed PDFs with the matrix element of only one initial state, and then to manually reweight the events by the ratio of the interference term to the original matrix element squared. Due to the potential issues with phase space sampling, this method should be validated on Standard Model processes, and is thus beyond the scope of the present work. }

Given these challenges, we do not attempt to implement the PDFs of the vector bosons in our calculations of vector boson fusion (VBF) production of the scalar mediators. 
Instead, we content ourselves with the effective vector approximation implemented in \texttt{MadGraph}~\cite{Ruiz:2021tdt}, which allows for a consistent treatment of initial state vector bosons. 
We use $\sqrt{\hat{s}}/2$ as the factorization scale for the PDFs, where $\hat{s}$ is the partonic squared center-of-momentum energy.

The other hurdles, relating to the helicities and the divergent $x \to 1$ behavior of the muon component of the PDFs, are more readily manageable with existing generators. The rest of this section is devoted to describing a pipeline for implementing the complete PDFs for the muon component of the muon beam in \texttt{MadGraph5}, accurately accounting for the modified kinematics and polarization effects at the cost of some additional processing of the generated events.

\subsection{Implementing the Muon Component of the PDFs}
\label{subsec:pdf_implementation}

We first separate the LHAPDF-style files from LePDF into separate files, each containing separate helicities of the muon/anti-muon. After removing the helicity information, the separated files can then be loaded into LHAPDF and interfaced with \texttt{MadGraph} in essentially the standard way. The helicity-split PDFs are then convolved with the matrix elements computed with the corresponding, specified helicity~\cite{BuarqueFranzosi:2019boy}. The resulting events from each process will have different weights, depending on both the couplings in the matrix element and the polarization effects in the muon PDFs themselves. So long as these weights are accounted for, the runs from different helicities can then simply be combined to obtain a complete sample from unpolarized muon beams.

Dealing with the divergent behavior as $x \to 1$ requires slightly more care. The integration over the parton luminosity must be cut off at some value of $x < 1$ for Monte Carlo integration to return a finite, reliable answer, but the singular region near $x \to 1$ is precisely where the bulk of the cross section lies.

To deal with this issue, we slice the phase space into regions with and without the PDFs in such a way that the numerical integration can be performed, and then stitch the results back together in a consistent way.\footnote{An alternative route to addressing this issue, implemented in the eMELA package in the context of the electron PDF in QED, is to augment the generator to replace the $x \to 1$ region with the known, analytic solution at some small value of $\epsilon = 1 - x$~\cite{Frixione:2019lga, Bertone:2019hks, Frixione:2012wtz, Bertone:2022ktl}. A similar methodology could be extended to the muon PDFs at higher energies, accounting for the helicity dependence in the exact solutions.}
To this end, we split the PDF of the muon into a ``continuum'' and a delta-function piece,
\begin{equation}
\label{eq:pdf_split}
f_{\mu}(x, Q^2) \simeq \tilde{f}_{\mu}(x, Q^2) + c_{\mu}(Q^2) \delta(1 - x), 
\end{equation}
where $\tilde{f}_{\mu}(x, Q^2)$ is identical to the original muon PDF (obtained, \textit{e.g.}, from the $(x, Q^2)$ grid provided by LePDF), multiplied by a smooth function that is unity for all $x \ll 1$ but rapidly approaches zero as $x \to 1$.\footnote{In practice, we use $\erf[(1-x)/\varepsilon]$, with $\varepsilon$ a parameter governing the width of the fall-off at $x \to 1$. We use values of $\varepsilon \in [10^{-2}, 10^{-8}]$ to assess the stability of our results, with $\varepsilon = 10^{-4}$ taken as the fiducial value.}

\begin{figure}
    \centering
    \includegraphics[width=0.65\linewidth]{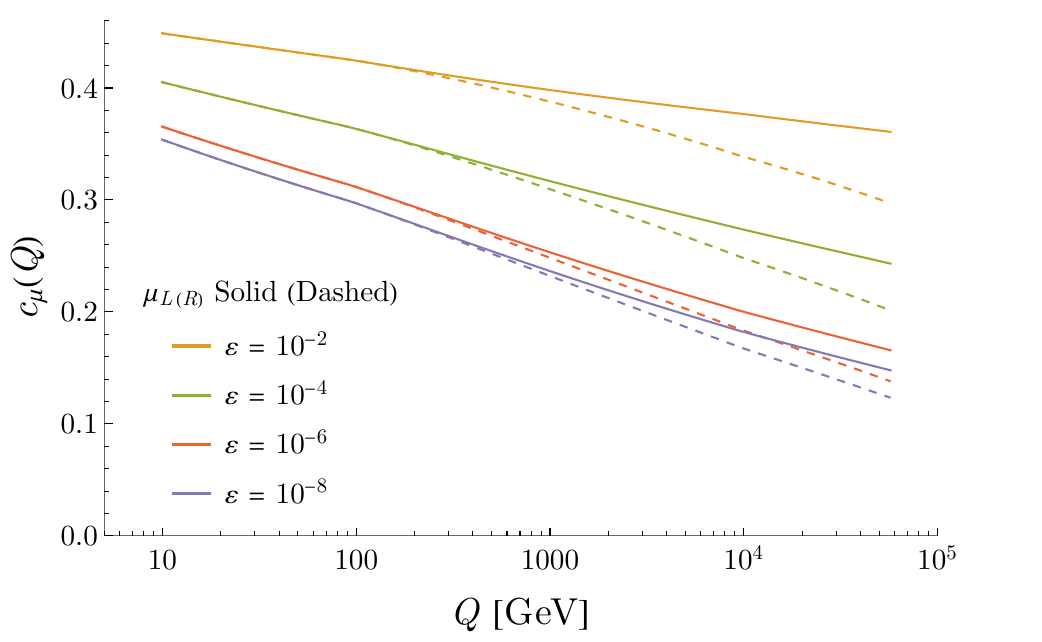}
    \vskip -0.2cm
    \caption{Values of the coefficients $c_{\mu_L, \mu_R}$ as a function of the factorization scale $Q$, for different values of the width $\varepsilon$ used in eliminating the $x \to 1$ limit of the muon component of the PDFs, as described in the text. The solid (dashed) lines show the values for the $L$ ($R$) helicities, respectively.}
    \label{fig:cLR_values}
\end{figure}

The prefactor $c_{\mu}(Q^2)$ in front of the delta-function is fixed by requiring the split PDF to satisfy the same sum rules as the original PDF. 
This splitting must be performed separately on the left- and right-helicity muon PDFs. In practice, we fix the values of $c_{\mu_L}$ and $c_{\mu_R}$ at each $Q^2$ by demanding that the total muon number, integrating over $x$ and summing over all muon and anti-muon components is unity, and that the total polarization (the ratio of left- to right-handed muon PDFs, separately integrated over $x$) is equal to the polarization computed with the original $f_{\mu}$. The resulting  values of $c_{\mu_L, \mu_R}$ depend on both $Q^2$ and on the precise manner in which $\tilde{f}_{\mu_L, \mu_R}$ approaches $0$ as $x \to 1$. This is demonstrated in \cref{fig:cLR_values}. We performed several cross-checks to ensure that these details do not meaningfully affect the results---the cross sections relevant for our models discussed in the next section varied at the percent level for different choices of the width used to eliminate the $x \to 1$ part of $f_{\mu}$.

After the splitting in Eq.~\eqref{eq:pdf_split}, the relevant parton luminosity can be decomposed into terms containing $\tilde{f}$ and terms containing $c_{\mu}$. The parton luminosity for two incoming muons is
\begin{equation}
\label{eq:parton_luminosity_decomp}
\begin{aligned}
\frac{\dd \mathcal{L}_{\mu^+\mu^-}}{\dd\tau}
& = \int_{\tau}^1 \frac{\dd x}{x} f_{\mu^+}(x,Q_f) f_{\mu^-}(\tau/x, Q_f) \\
& = \frac{\dd \tilde{\mathcal{L}}_{\mu^+\mu^-}}{\dd\tau} 
+ c_{\mu^+} c_{\mu^-} \delta(1 - \tau)
+ c_{\mu^+} \tilde{f}_{\mu^-}(\tau) + c_{\mu^-}\tilde{f}_{\mu^+}(\tau),
\end{aligned}
\end{equation}
where $\tau = \hat{s} / s$ is the fraction of the total center of mass, $Q_f$ is the chosen factorization scale (at which all functions are evaluated in the second line), and $\tilde{\mathcal{L}}$ is the parton luminosity computed with both $f \to \tilde{f}$.
Each of the four terms in the second line of Eq.~\eqref{eq:parton_luminosity_decomp} can be readily computed in \texttt{MadGraph5} with the $\tilde{f}_{\mu}$ interfaced via $\textsc{LHAPDF}$. The first term is computed with both initial state muons drawn from the truncated PDFs, the second term with both muons taken to have the full energy of the beam (no PDFs), and the final two terms with one muon taken with the full beam energy and the other with the PDF. The corrected cross section along with samples of (weighted) events are obtained by adding the samples generated from these four processes, multiplied by the appropriate factors of $c_{\mu}(Q_f)$, computed independently. The divergences do not appear due to the fall off in the $\tilde{f}$ as $x \to 1$.

\subsection{Effects of the Muon Parton PDFs}
\label{subsec:pdf_effecs}

We now consider the effects of including the PDFs for the muon, compared to the leading order calculation. We will focus on the $e$ model, to facilitate comparison with the results of Ref.~\cite{Asadi:2023csb}.
The effects of including the PDFs for the muon component are quite similar for all the benchmark fermion-portal models that we considered. 

\begin{figure}
    \centering
    \includegraphics[width=0.95\linewidth]{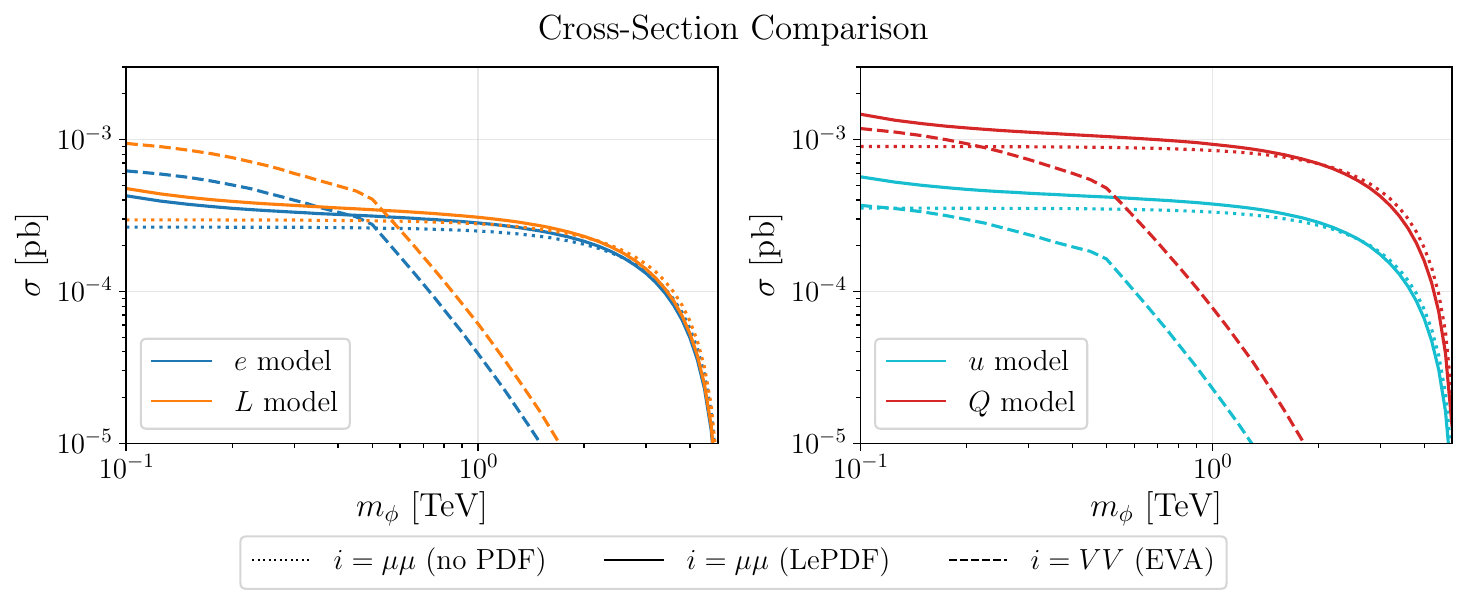}
    \caption{Cross section for mediator pair production for lepton-portal (\textbf{left}) or quark-portal (\textbf{right}) models. The solid (dashed) lines show muon (VBF) initial states, including the muon parton PDF (EVA). The dotted lines shows muon initial states with no PDF. Note that the kink in the $VV$ initial state curve is due to a cutoff on the partonic center-of-momentum energy $\hat{s}$ at 1 TeV, which is necessary to get accurate results from EVA (see Ref.~\cite{Ruiz:2021tdt}). We find that including the muon parton PDF only affects the total cross section at low mediator masses. We will use these cross sections in our collider study (\cref{sec:results}).}
    \label{fig:ellR_xsec}
\end{figure}

In Fig.~\ref{fig:ellR_xsec}, we show the cross section for $\phi \phi$ pair production, with various initial states, with the muon initial states shown with and without the PDFs. We see that including the muon PDF leads to a slight increase in the cross section, except for $\phi$ masses approaching the beam energy.
This can be attributed to the fact that at large $\phi$ masses, a smaller fraction of the muons will carry enough energy in the collision to exceed the $2m_{\phi}$ threshold. As the $\phi$ mass becomes smaller, however, this effect becomes less significant, and instead we see an enhancement from the fact that the parton-level cross sections are larger at smaller center-of-momentum energies.

\begin{figure}
    \centering
    \includegraphics[width=0.48\linewidth]{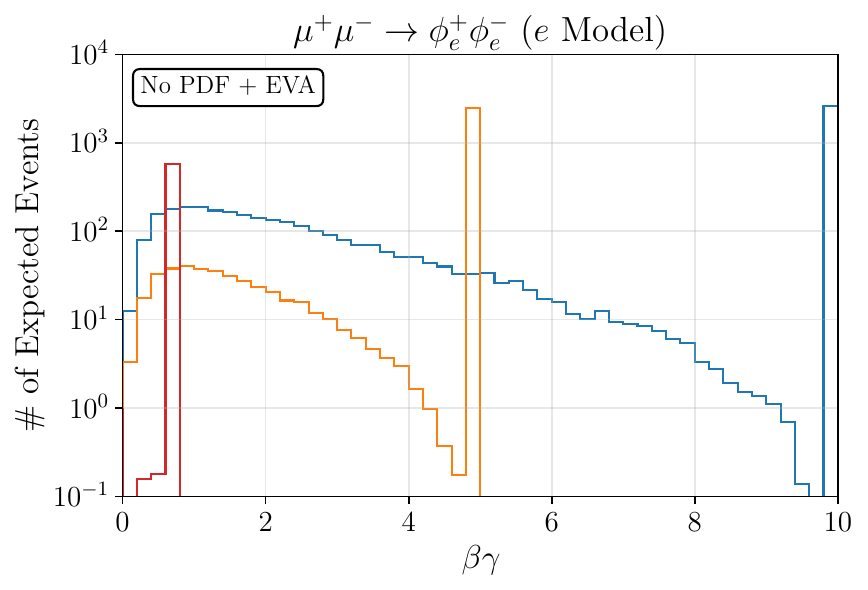}
    \includegraphics[width=0.48\linewidth]{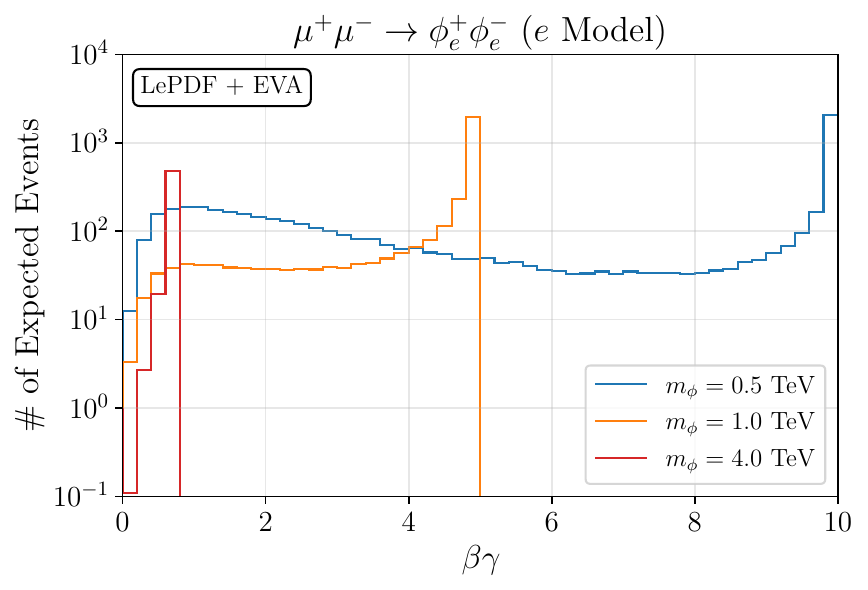}
    \caption{\underline{$e$ model}: Histograms of $\beta \gamma$ of the $\phi$ final states. The \textbf{left} (\textbf{right}) figure shows the distribution of $\beta \gamma$ without (with) muon parton PDFs. The colors correspond to three different mass points ({\bf\textcolor{matBlue}{blue}} = 500 GeV, {\bf\textcolor{matOrange}{orange}} = 1 TeV, {\bf\textcolor{matRed}{red}} = 4 TeV). As expected, including the muon component of PDFs smoothens the sharp peak at large values of $\beta \gamma$, see the text for more details. }
    \label{fig:ellR_LLP_hist}
\end{figure}

While the total cross section is not affected significantly, various kinematic distributions go through a bigger change when the muon PDF is included. 
Figure~\ref{fig:ellR_LLP_hist} shows the effect of this PDF on the kinematic variable $\beta\gamma$, which was explored in Ref.~\cite{Asadi:2023csb}.
As in the prior study, we see that there is still a peak at the largest value, where the muon component of the PDFs carry essentially all of the beam energy, however the peak is slightly smaller and there is a more gradual decrease in the next highest bins. Instead, the distribution falls smoothly until the VBF production mode becomes dominant, mirroring the shape of the PDFs themselves.
Overall, we find that the inclusion of the muon PDF has a non-negligible effect on the kinematic distributions.

\begin{figure}
    \centering
    \includegraphics[width=0.6\linewidth]{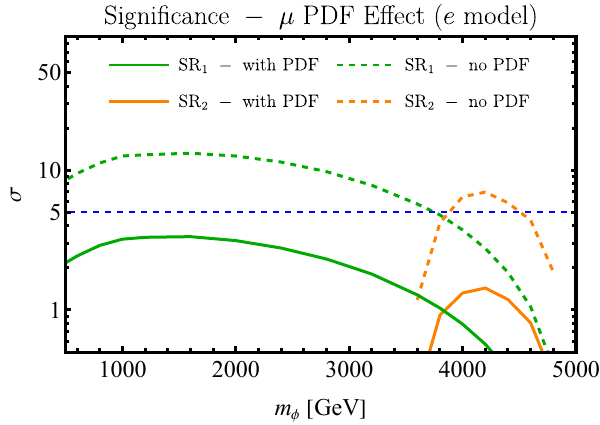}
    \caption{\underline{$e$ model}: Achievable discovery significance of the $e$ model for two signal regions (SRs) used in Ref.~\cite{Asadi:2023csb}. Solid (dashed) lines show the result with(out) the muon component of the PDFs. Including this PDF drastically diminishes the efficacy of these signal regions to below the discovery threshold (dashed blue line) for the entire mass range. }
    \label{fig:sig_drop}
\end{figure}

As a result, cuts designed to exclude the background are less efficient at preserving signal events, as the affected variables are precisely among those used to distinguish the signal from the background. This results in a significant decrease in the reach, compared to the analysis without the muon parton PDFs. This effect is illustrated in~\cref{fig:sig_drop}, which shows the achievable significance as a function of $m_{\phi}$ for two of the signal regions from Ref.~\cite{Asadi:2023csb}. 
This figure underscores the importance of including the muon PDF in a proper study of fermion-portal models, and possibly other BSM scenarios, at a future muon collider.

\begin{figure}
    \centering
    \includegraphics[width=0.9\linewidth]{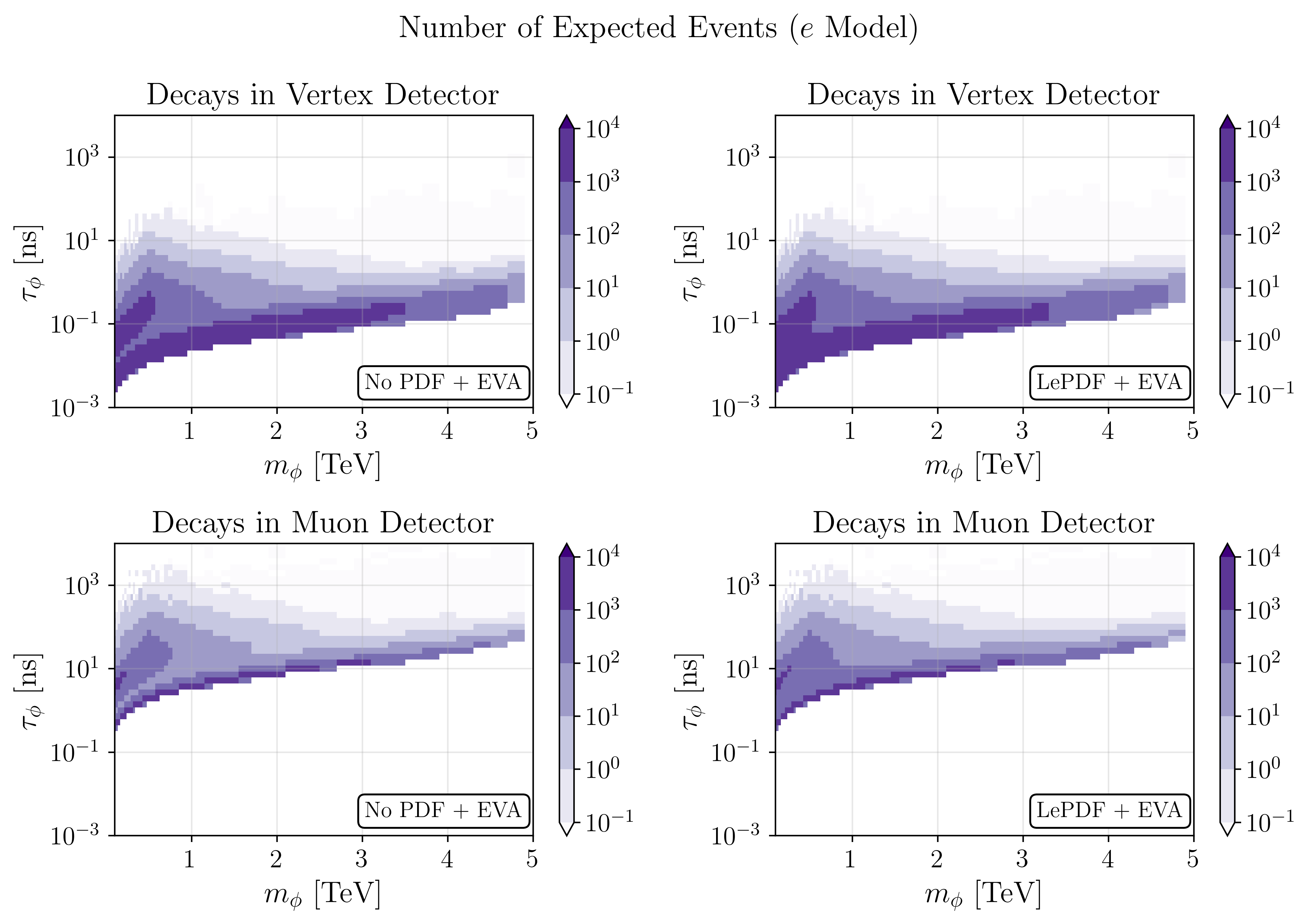}
    \caption{\underline{$e$ model}: Number of expected displaced leptons in two regions of the detector for a 10 TeV MuC, assuming the tentative design of Table~\ref{tab:detector}. The {\bf left} ({\bf right}) plots correspond to the prediction without (with) the muon parton PDF. The DM Yukawa coupling is chosen such that it reproduces the correct relic abundance today. In both detector regions, including the muon PDF fills up the region between the two cluster of events. This is attributed to the more gradual growth of distributions between the two peaks in various kinematic distributions, see \cref{fig:ellR_LLP_hist}. }
    \label{fig:eR_LLP_diff}
\end{figure}

These conclusions carry over to the long-lived regime, where the effect of the PDFs is essentially to populate the gap between the highly-dense regions arising from initial state muons or VBF. We illustrate this in~\cref{fig:eR_LLP_diff}, showing the number of expected events in the $m_{\phi}$ vs. $\tau_{\phi}$ plane for different parts of the detector. 
All of these results together motivate further work on implementing the muon PDFs more directly in event generators to facilitate more reliable predictions of the reach for various models. 

\section{Muon Collider Phenomenology}\label{sec:results}

In this section, we study the signals of each fermion-portal model at a future muon collider, folding in the effects of the muon PDF as described in the previous section. As in the previous section, we will study a 10 TeV muon collider with a target luminosity of 10~ab$^{-1}$ \cite{Delahaye:2019omf}. 
In what follows, we will assume that the DM $\chi$ populates the entire DM relic abundance today, which fixes its couplings $\lambda$ to the mediators (see \cref{fig:lambda_L,fig:lambda_lifetime_u,fig:lambda_Q}). 
To streamline our calculation, we will also assume the mediators interact with all SM fermion generations with the same coupling and will remain (mostly) inclusive over the final state SM fermions in our signals.

We study the pair production signals of mediators for each model. 
In Section~\ref{sec:models}, we saw that all models can give rise to either prompt or LLP signals, depending on the parameters of the model. 
The threshold between a prompt or an LLP signal depends on the details of the search, the model, and the detector design. For our purposes, we define ``prompt" signals to be those where the charged mediator has $c\tau_\phi \lesssim 1$~cm. 

\begin{figure}
    \centering
    \resizebox{\columnwidth}{!}{
    \includegraphics[width=0.5\linewidth]{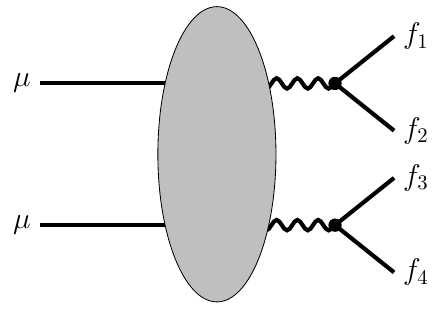}
    \includegraphics[width=0.5\linewidth]{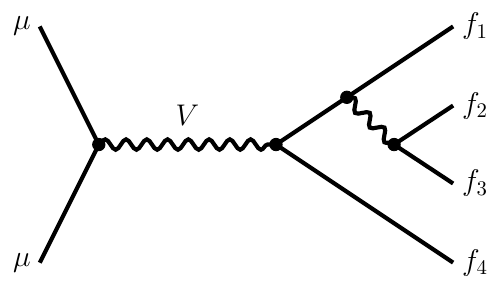}
    \includegraphics[width=0.5\linewidth]{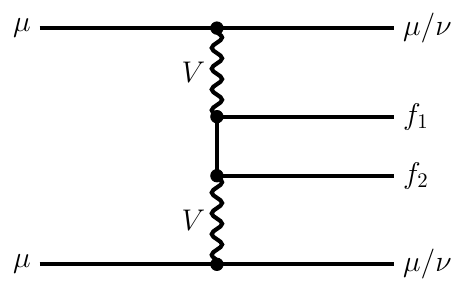}
    }
    \caption{Background topologies at a future muon collider for all fermion-portal models under study. They include a pair production of electroweak gauge bosons $Z$ or $W$ and their subsequent decay to SM fermions (\textbf{left}), a muon annihilation process with an FSR (\textbf{middle}) or an ISR (not shown), and VBF production of SM fermions (\textbf{right}). These backgrounds can be subtracted by cuts on invariant mass of visible SM fermions and $M_{T2}$, their opening angle, and their energies, respectively; see the text for more details. }
    \label{fig:bkgd_diagrams}
\end{figure}

In the prompt region, we simulate backgrounds and put forward simple cut-and-count analyses. The primary backgrounds for our collider searches are $Z$- and $W$-pair production, initial and final state radiation (ISR and FSR), and a VBF background, as shown in Fig.~\ref{fig:bkgd_diagrams}. 
We use the following kinematic cuts to distinguish between our fermion-portal DM signal and these backgrounds: 
\begin{itemize}
    \item $m_{\ell\ell,jj}$ -- invariant mass of a lepton or a jet pair: A cut above the $Z$ pole on the invariant mass distribution of the lepton pair (for the lepton-portal models) or jets (for the quark-portal models) alleviates the $Z$-pair production background.
    \item $M_{T2}$ -- transverse mass for semi-invisible decays~\cite{Lester:1999tx}: For the $W$-pair production background, this variable is bounded from above by the $W$ mass. Therefore, a cut on $M_{T2}$ removes the $W$ background. However, the exact value of the cut depends on the signal parameters. As we will discuss in the upcoming sections, events from other background processes give rise to a monotonically decreasing $M_{T2}$ distribution, whereas the signal distribution will be uniform, see \cref{fig:prompt_histo_e}. Thus, cuts on this observable help in subtracting other backgrounds as well.
    \item $\theta_{\ell\ell,jj}$ -- opening angle in the lab frame between $\ell^+\ell^-$ or $jj$: In the signal events, and especially for light mediators that are boosted, the leptons/jets will be produced back-to-back and have a large opening angle. The ISR and FSR backgrounds, instead, will give rise to a smaller opening angle between the leptons/jets. Therefore, a cut on $\theta$ will help remove these backgrounds. 
    \item $E_{\ell,j}$ --  the lepton or jet energy: The VBF background will produce lower energy leptons/quarks (for the lepton--portal and quark--portal models, respectively), thus a lower bound on $E_{\ell,j}$ will suppress the VBF background. 
    \item $\eta_{\ell,j}$ -- 
    pseudorapidity: As we will see in the upcoming section (see \cref{fig:prompt_histo_e}), the signal events are more central than the background, motivating an upper bound on $\eta_{\ell,j}$. 
\end{itemize}

\begin{table}
    \centering
    \resizebox{\columnwidth}{!}{
    \begin{tabular}{r|l||c|c|c|l}
        \textbf{Subsystem} & \textbf{Region} & \textbf{$L$ dimensions [cm]} & \textbf{$|Z|$ dimensions [cm]} & $\mathbf{\eta}$ \textbf{bound} & \textbf{Material} \\
        \hline\hline
        Vertex Detector & Barrel & $3.0 - 10.4$     & $65.0$    & $\lesssim 2.53$ 
        & Si \\
                        & Endcap & $2.5 - 11.2$     & $8.0 - 28.2$  & $\lesssim 1.65$ 
                        & Si \\
        \hline
        Inner Tracker   & Barrel & $12.7 - 55.4$    & $48.2 - 69.2$  & $\lesssim 1.05$ 
        & Si \\
                        & Endcap & $40.5 - 55.5$    & $52.4 - 219.0$  & $\lesssim 1.07$ 
                        & Si \\
        \hline
        Outer Tracker   & Barrel & $81.9 - 148.6$   & $124.9$     & $\lesssim 0.76$ 
        & Si \\
                        & Endcap & $61.8 - 143.0$   & $131.0 - 219.0$ & $\lesssim 1.21$ 
                        & Si \\
        \hline\hline
        ECAL            & Barrel & $150.0 - 170.2$  & $221.0$      & $\lesssim 1.08$ 
        & W + Si \\
                        & Endcap & $31.0 - 170.0$   & $230.7 - 250.9$ & $\lesssim 1.18$ 
                        & W + Si \\
        \hline
        HCAL            & Barrel & $174.0 - 333.0$  & $221.0$      & $\lesssim 0.62$ 
        & Fe + PS \\
                        & Endcap & $307.0 - 324.6$  & $235.4 - 412.9$ & $\lesssim 0.71$ 
                        & Fe + PS \\
        \hline\hline
        Solenoid        & Barrel & $348.3 - 429.0$  & $412.9$        & $\lesssim 0.85$ 
        & Al \\
        \hline\hline
        Muon Detector   & Barrel & $446.1 - 645.0$  & $417.9$     &   $\lesssim 0.61$ 
        & Fe + RPC \\
                        & Endcap & $57.5 - 645.0$   & $417.9 - 563.8$  & $\lesssim 0.79$ 
                        & Fe + RPC \\
    \end{tabular}
    }
    \caption{Different components of the current proposed detector design for a high energy MuC. Table is adapted from Ref.~\cite{MuonCollider:2022ded} with the addition of the $\eta$ bound column, which shows the highest value of the pseudorapidity $\eta$ for a track that completely goes through that region. The $L$ ($Z$) dimension refers to the segment size in the transverse (longitudinal) direction. This information enters our LLP search.}
    \label{tab:detector}
\end{table}

For the LLP section, we report the signal count for each part of the detector as described in Table~\ref{tab:detector}, and is modeled after~\cite{MuonCollider:2022ded,Black:2022cth}, which is in turn informed by the existing designs for CLIC \cite{CLICdp:2017vju} and ILC \cite{ILC:2007vrf}. 
Here, we do not perform a cut-and-count analysis as the irreducible background for the LLP searches arises from the detector response and requires detailed detector simulation, e.g., with \textsc{Geant4}, which is beyond the scope of this work. 
Similar to the analysis of Ref.~\cite{Asadi:2023csb}, we use the information from the barrel regions of detectors as the $dE/dx$ measurements are slightly better in these regions and there are fewer high mass background tracks in the barrel regions compared to endcaps.
For all models, we also include the current LHC bounds.

For the lepton- (quark-) portal models, the main LLP signatures are visible tracks that end at a displaced lepton (vertex), see Ref.~\cite{Asadi:2023csb} for the calculation of the transverse displacements of the track before its decay.  
In the lepton-portal models, each signal event consists of a charged track, while for the quark--portal models, where the mediator is color-charged, the signal event is a jet ($R$-hadron \cite{Farrar:1978xj,Arkani-Hamed:2004ymt} since the color-charged particle is long-lived). 
As discussed before, the two production channels (muon annihilation and VBF---see \cref{tab:diagrams}) give rise to different kinematic distributions such that displaced vertices arising from these channels appear in different parts of the detector.

The background from the SM processes can be reduced with the help of various observables, thus the main irreducible background for LLP searches are from detector response.
In particular, the time-of-flight and $dE/dx$ will be different than for SM particles owing to $\phi$'s large mass. See Ref.~\cite{Asadi:2023csb} for a study of time-of-flight in the $e$ model.

For the $e$  and $u$ models, there is only one mediator in the spectrum with only one decay channel. This gives rise to similar lifetimes for these models on their mediator and DM mass planes, see \cref{fig:lambda_parameter,fig:lambda_lifetime_u}. 
We will find that for a large part of their parameter space, the mediator in these two models is long-lived enough to give rise to a heavy stable charged particle. 
On the other hand, the $L$  and the $Q$ models have two mediators in the spectrum, the heavier of which can decay via electroweak interactions, resulting in an upper bound on their lifetime. Thus, these mediators decay inside the detector and give rise to displaced vertices in different detector components for most of their parameter space. 
In the $L$ model case the lighter daughter mediator is invisible, giving rise to a disappearing track signal. 
On the other hand, in the $Q$ model the lighter daughter mediator is still detectable and appears as a continuation of the original $R$-hadron. Thus, in this model we will study the location of the subsequent lighter mediator's decay; the signal in this case will be an $R$-hadron that either eventually decays to a jet and missing energy, or leaves the detector.

\subsection{$e$ model}
\label{subsec:collidere}

In this section, we update the studies of the $e$ model from Ref.~\cite{Asadi:2023csb} to include the effects of the muon PDFs. As discussed in \cref{sec:PDF}, the muon PDF affects the kinematic distributions and therefore will modify the optimal kinematic cuts. In the prompt region, we find that including the muon PDFs diminishes the mass reach of a future muon collider to the point that with the set of cuts proposed in Ref.~\cite{Asadi:2023csb}, we cannot discover the mediator for any masses beyond the current LHC bounds. 
We will show instead that cuts on observables discussed above will allow us to discover this model as long as $m_\phi \lesssim 3.3$~TeV. 
This should be contrasted to bounds from recasting current LHC slepton searches \cite{ATLAS:2019lff,CMS:2020bfa}, which only probe the model for mediator masses $m_\phi \lesssim 0.7$~TeV. 
On the LLP part of the parameter space, including the PDF distorts the distribution of final state events enough to fill in some area between the observed ``double peak'' feature, which arises from the two production channels (muon annihilation or VBF). 

While the branching ratios to final SM particles will vary from one model to another, the signal and background diagram topologies of other fermion portals are identical to this model, which leads to similar event counts and kinematic distributions. 
As a result, we will provide more details in this section and repeat the analysis of this model for other portals as well, in both the prompt and the LLP region.

\subsubsection{Prompt Region}
\label{subsubsec:prompt_e}

The prompt signal for the $e$ model is from the process $\mu^+\mu^-\to \phi_e^+\phi_e^-$ with the subsequent decay $\phi_e^\pm\to \ell^\pm\chi$, which results in a pair of charged leptons and missing energy. Since we assume a democratic coupling to all lepton flavors, the branching ratio of $\phi_e\to \ell_i\chi$ is 1/3 for each lepton flavor $i$. 
To avoid dealing with complications of reconstructing a tau lepton, and to maximize the signal count, we focus on events with $\ell =e,\mu$ in the final state.

The background topologies are included in \cref{fig:bkgd_diagrams} and arise from the production of $W$s or $Z$s with subsequent leptonic decays, ISR or FSR of one of these gauge bosons, or a pair production of leptons via a vector boson fusion.\footnote{Similar to the analysis of Ref.~\cite{Asadi:2023csb}, there are backgrounds arising from VBF initial states that can be approximated by the Improved Weizsacker--Williams (IWW) method; they can be shown to contribute negligibly to the total background and can be safely ignored. }
While these have very similar topologies to the signal (especially the $W$ pair production diagram), there are strong kinematic cuts that can distinguish the signal, as discussed in the introduction to this section. 
Specifically, 
we find that the $Z$-pair production background can be removed by
a cut on the invariant mass of $m_{\ell\ell} \geqslant 300$~GeV. 
The choice of the remaining cuts ($\theta_{\ell\ell}, M_{T2},E_\ell,\eta_\ell$) can be understood by looking at the distributions of events in terms of the remaining kinematic variables.

\begin{figure}[t]
            \centering
            \resizebox{\columnwidth}{!}{
            \includegraphics[width=0.8\linewidth]{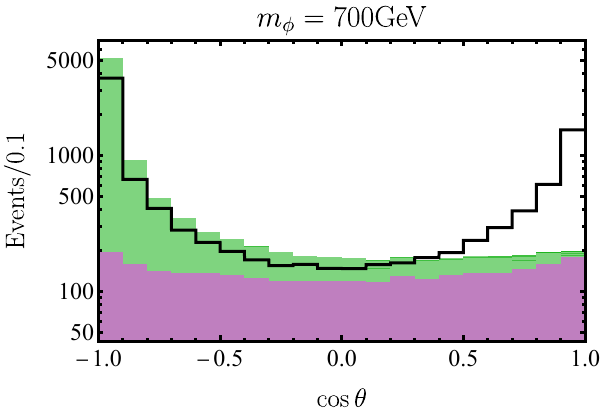}
            \includegraphics[width=0.8\linewidth]{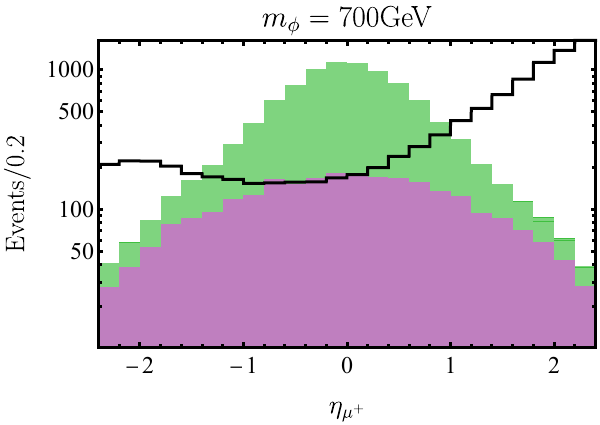}
            \includegraphics[width=0.8\linewidth]{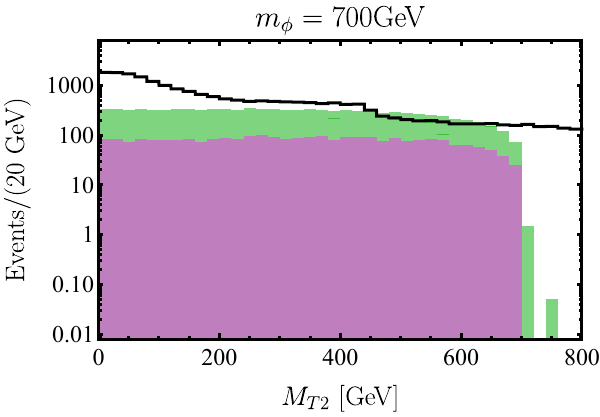}
            }\\
            \resizebox{\columnwidth}{!}{
            \includegraphics[width=0.8\linewidth]{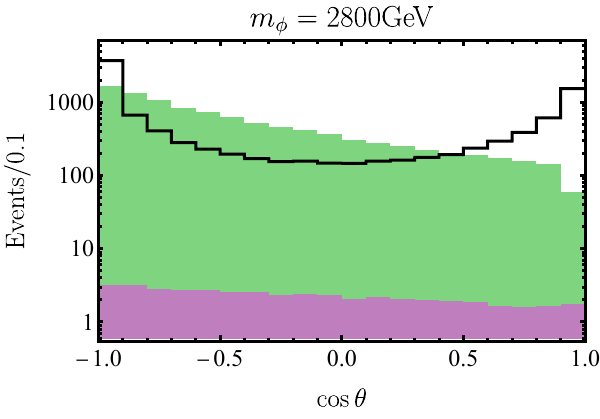}
            \includegraphics[width=0.8\linewidth]{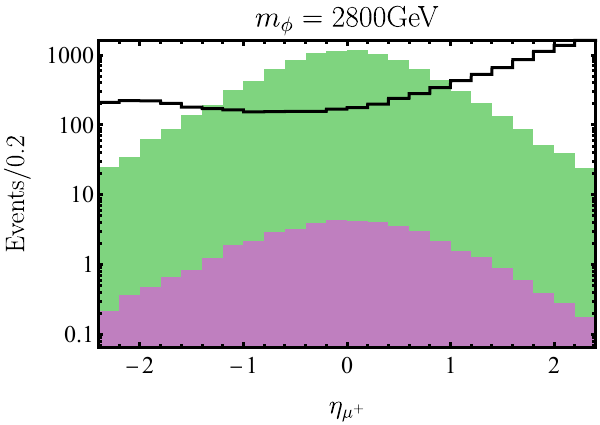}
            \includegraphics[width=0.8\linewidth]{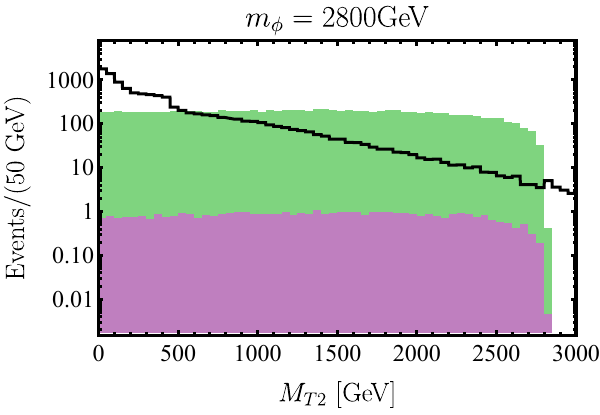}
            }
            \caption{\underline{$e$ model:} Distribution of events for the opening angle between charge leptons in the final state (\textbf{left columns}), rapidity of the final leptons (\textbf{middle columns}), and $M_{T2}$ (\textbf{right columns}) for two benchmark mediator masses 0.7~TeV (\textbf{top row}) and 2.8~TeV (\textbf{bottom row}). The signal events from initial VBF are denoted by {\bf \color{PouyaPurple}purple}, the signal events from initial muons are denoted by {\bf \color{PouyaGreen}green}, and the {\bf black} line denotes the background distribution. 
            Both the background and the total signal histograms are made with ten thousand events; the contribution of each channel to the signal histograms are weighted by their cross sections. In addition to these three observables, we use cuts on $m_{\ell\ell}$ and on $E_\ell$ to suppress the background, see the text for further details.}
            \label{fig:prompt_histo_e}
\end{figure}

To show the power of different kinematic cuts, and for two benchmark masses of the mediator, in \cref{fig:prompt_histo_e} we show the distribution of events in some of these observables. 
We separate the contribution of different production channels for the signal, weighted proportional to their cross sections. 
These channels include the VBF production of the mediator pair and the contribution from initial state muons. 
In our implementation of muon PDF, the latter includes four possibilities depending on whether the muon from each beam enters the collision with $x\rightarrow 1$ or not, see \cref{eq:parton_luminosity_decomp}.
For low mediator masses the VBF is the dominant production channel, and as we go to higher masses more events arise from the initial muon channel, as expected from \cref{fig:ellR_xsec}.

We clearly see that the background has a different distribution than the signal channels in these kinematic variables. 
The peak in the background distribution in $\cos\theta \rightarrow -1$ arises from $W$-pair production events since the two $W$s are very boosted and move back-to-back when the incoming energy from the two beams are equal. 
This background can, in principle, be discarded with a cut on the $M_{T2}$ observable since for these events $M_{T2} \leqslant m_W$ \cite{Lester:1999tx}.

We find that the following cuts: 
\begin{align}
    \label{eq:SR_e_def}
    \mathrm{SR}_1^\ell(m_\phi=0.7~\rm{TeV})&: \left\lbrace m_{\ell\ell} \geqslant 300~\mathrm{GeV}, \theta_{\ell\ell} \geqslant \frac{8\pi}{9},|\eta_{\ell}| \leqslant 0.3, M_{T2} \geqslant 150~\mathrm{GeV}, E_{\ell} \geqslant 600~\mathrm{GeV}  \right\rbrace , \\
    \mathrm{SR}_2^\ell(m_\phi=2.0~\rm{TeV})&: \left\lbrace  m_{\ell\ell} \geqslant 300~\mathrm{GeV}, \theta_{\ell\ell} \geqslant \frac{13\pi}{18},|\eta_{\ell}| \leqslant 0.2, M_{T2} \geqslant 800~\mathrm{GeV}, E_{\ell} \geqslant 1800~\mathrm{GeV}  \right\rbrace  ,
    \nonumber 
\end{align}
maximize the significance of the signal for $m_\phi=0.7$~TeV  and $m_\phi=2$~TeV, respectively. We see that heavier $\phi$ requires more stringent (larger) $M_{T2}$ and $E_\ell$ cuts, and have smaller $\theta_{\ell\ell}$ and $\eta_\ell$, as expected. 
We will use these two signal regions in our search.

\begin{figure}
            \centering
            \resizebox{\columnwidth}{!}{
            \includegraphics[width=0.6\linewidth]{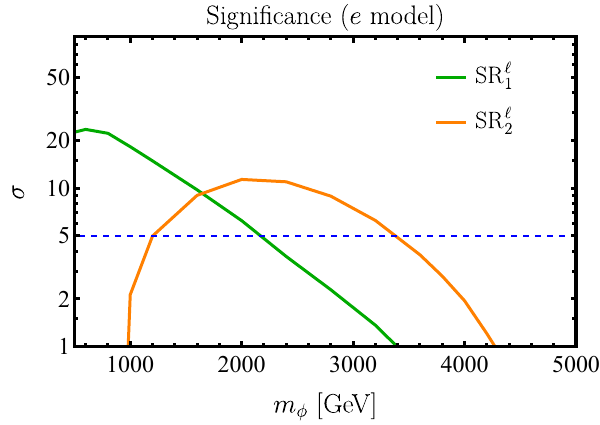}
            \hspace{0.1in}
            \includegraphics[width=0.62\linewidth]{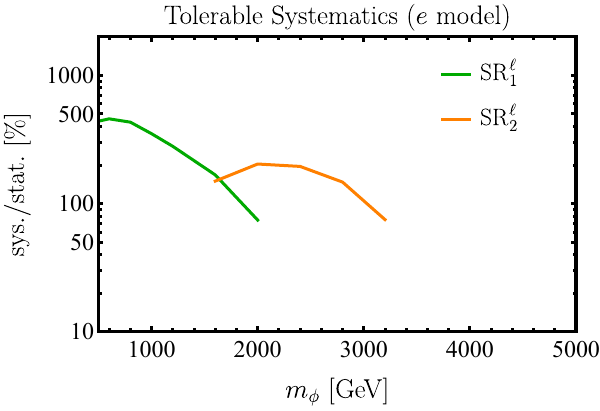}
            }
            \caption{\underline{$e$ model:} The best achievable discovery significance with zero systematic uncertainty (\textbf{left}) and the tolerable uncertainty that still allows a $5\sigma$ discovery of the mediator (\textbf{right}) for the signal regions defined in Eq.~\eqref{eq:SR_e_def}. The dashed blue line on the left plot marks the $5\sigma$ discovery threshold. We find that even with systematic uncertainties as large as the statistical ones the model is discoverable in a future muon collider for mediator masses up to $m_\phi \lesssim 3.3$~TeV. }
            \label{fig:prompt_result_e}
\end{figure}

In \cref{fig:prompt_result_e}, we show the discovery reach of our proposed search in the parameter space of the $e$ model. 
We find that even with only the two signal regions in Eq.~\eqref{eq:SR_e_def}, we can still discover the model (with $5\sigma$ significance) for mediator masses $m_\phi \lesssim 3.3$~TeV. 
This is far below the reach of $m_\phi\lesssim 4.5$ TeV calculated in Ref.~\cite{Asadi:2023csb} and, as discussed in the previous section, the drop is attributed to the effect of non-trivial muon PDF on the event distribution.

So far we have not included any systematic uncertainties in our analysis. Instead, in \cref{fig:prompt_result_e} we show the tolerable systematics (as a fraction of statistical uncertainties) that still allow us to discover this model for different mediator masses in the prompt region. 
It should be noted that ours is merely a rudimentary analysis with focus on a limited set of kinematic variables. More sophisticated studies of the kinematic distributions, as well as use of suitable machine learning techniques can enhance the reach of a future muon collider in our parameter space.
Given the theoretical appeal of this minimal DM setup, we hope our results on the tolerable systematics can inform the on-going efforts on designing a detector for a muon collider.

\subsubsection{Long-Lived Particle Region}
\label{subsubsec:LLP_e}

In \cref{fig:eR_LLP_diff} we see the effect of the muon PDF on the number of expected events in different detector components on the $\tau_\phi - m_\phi$ plane. Although there is still a double peak feature, which arises from the two production channels initial muons vs. VBF, visible in the distribution of $\beta \gamma$, the the gap that existed between the two peaks has been filled in as compared to the distributions without the PDF.

\begin{figure}
    \centering
    \includegraphics[width=0.9\linewidth]{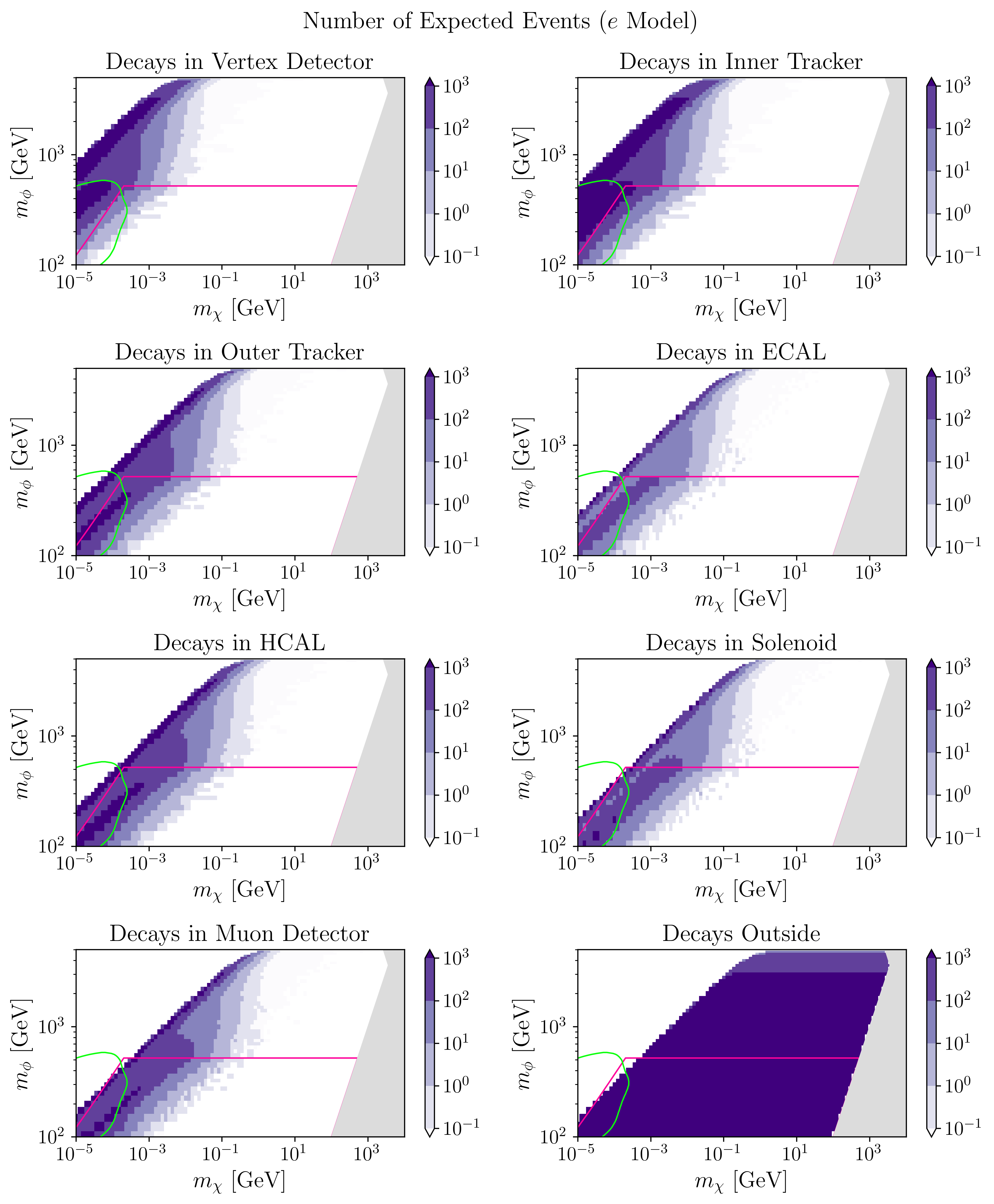}
    \caption{\underline{$e$ model:} Number of expected displaced leptons in different barrel regions of the detector, assuming the tentative design of Table~\ref{tab:detector}, as well as number of expected detector-stable charged tracks.  
    The gray region corresponds to $m_\chi > m_\phi$ (lower half) or DM overclosing the universe (upper half) and is not phenomenologically viable. For the entire LLP region of the parameter space, we find $\sim 10^3$ events (either a kinked charged track or a heavy stable charged track) in some part of the detector. The {\bf\textcolor{xkcdNeonGreen}{green}} and {\bf\textcolor{xkcdNeonPink}{pink}} regions indicate current LHC slepton constraints from Ref.~\cite{ATLAS:2020wjh} and Ref.~\cite{CMS:2024qys}, respectively. }
    \label{fig:eR_LLP_mchi}
\end{figure}

Figure~\ref{fig:eR_LLP_mchi} shows the number of expected events in each detector region on the $m_\phi - m_\chi$ plane, including the effects of the muon PDF. 
Each event consists of a long-lived charged track that eventually decays to a lepton and missing energy via the freeze-in coupling. 
In the figure, we report the number of these displaced decay vertices in each detector component. 
It should be noted that in this figure, as well as counterpart figures for other models, we report the average event count in each detector component, while the actual count will be drawn from a Poisson distribution whose parameter is given by our calculation in~\cref{fig:eR_LLP_mchi}. 
For most of the parameter space the mediator will be a heavy stable charged particle, while for low DM masses the freeze-in coupling is large enough that the mediator can decay inside the detector.
The conclusion is that we expect $>10^3$ events in almost the entire parameter space.

\subsection{$L$ model}
\label{subsec:colliderL}

In this section we study the collider signatures of the $L$ model. 
Unlike the $e$ model studied above, in this setup we have two nearly mass-degenerate mediators.
This has two major implications for the collider signals of this setup: (i) it provides a new 3-body decay channel for the charged mediator via the weak interaction, which puts an upper bound on its lifetime (see~\cref{fig:lifetime_L}), and (ii) thanks to the small mass splitting, when the charged mediator decays to the neutral mediator and charged SM leptons, the latter will be very soft, giving rise to a disappearing track signal in the LLP region.

The neutral mediator in this setup decays predominantly invisibly (to a DM particle and a neutrino).\footnote{As shown in \cref{tab:diagrams}, this mediator can decay to four final state fermions via an off-shell $W$ and an off-shell charged mediator, giving rise to hard final state SM fermions. However, the branching ratio to this channel is suppressed by the small freeze-in coupling and we find that it is sub-percent for almost the entire parameter space. As a result of this, the rate for this signal is very sub-dominant compared to other decay channels and is not further studied in this work.} 
As a result, and in line with typical jet+MET searches for DM at LHC \cite{ATLAS:2021kxv,CMS:2021snz,ATLAS:2024kpy,CMS:2024zqs}, one should rely on ISR/FSR or forward detectors to tag these events \cite{Han:2020uak,Ruhdorfer:2024dgz}. 
While the correlation between this signal and the charged mediator signals can be used to distinguish this model from other BSM signals, in our rudimentary analysis below we will only focus on searching for signals of the charged mediator and merely note that the neutral mediator can be searched for in, \textit{e.g.}, single photon searches at a future muon collider. 
Current LHC slepton searches \cite{ATLAS:2019lff,CMS:2020bfa} can probe this model for mediator masses $m_\phi \lesssim 0.7$~TeV.

\subsubsection{Prompt Region}
\label{subsubsec:prompt_L}

The two available decay channels for the charged mediator are (i) via the electroweak interaction to a very soft lepton and missing energy, \textit{i.e.} invisible decay, or (ii) to DM and a charged lepton via the $\lambda$ coupling. 
The former decay channel can be searched for using potential initial state radiation (similar to the neutral mediator), while the decay to DM and a charge lepton gives rise to a clear lepton+MET signal. 
We will use the latter channel in our search as it is the dominant decay channel for the charged mediator in the prompt region of its parameter space, see \cref{fig:BR_L}.
The background for this search is identical to the search for the $e$ model of the previous section, see \ref{subsec:collidere} for a discussion of the shape of the background distribution. 

\begin{figure}
            \centering
            \resizebox{\columnwidth}{!}{
            \includegraphics[width=0.6\linewidth]{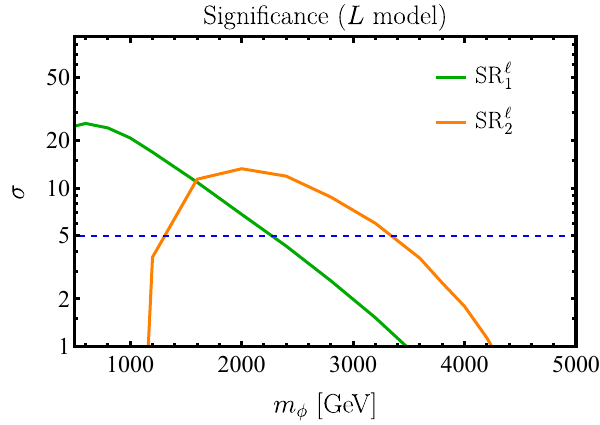}
            \hspace{0.1in}
            \includegraphics[width=0.62\linewidth]{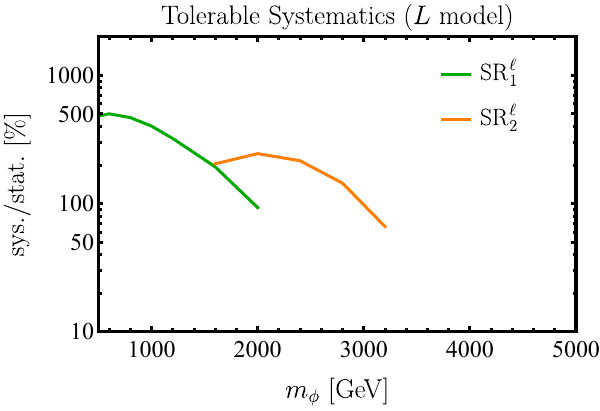}
            }
            \caption{\underline{$L$ model:} The best achievable discovery significance with zero systematic uncertainty (\textbf{left}) and the tolerable uncertainty that still allows a $5\sigma$ discovery of the mediator (\textbf{right}) for the signal regions defined in Eq.~\eqref{eq:SR_e_def}. The dashed blue line on the left plot marks the $5\sigma$ discovery threshold. We find that even with systematic uncertainties as large as the statistical ones the model is discoverable in a future muon collider for mediator masses up to $m_\phi \lesssim 3.4$~TeV. }
            \label{fig:prompt_result_L}
\end{figure}

We find that the cuts introduced in Eq.~\eqref{eq:SR_e_def} for the $e$ model can also extract the signal from the background in this model very efficiently. 
We show the reach of our proposed search in the parameter space of this model in \cref{fig:prompt_result_L}.
Using these signal regions, and neglecting the systematic uncertainties, we can discover this model for mediator masses $m_\phi \lesssim 3.4$~TeV, a significant improvement compared to current bounds of $m_\phi \lesssim 0.7$~TeV from LHC's slepton searches \cite{ATLAS:2019lff,CMS:2020bfa}.

In \cref{fig:prompt_result_L} we also show the tolerable systematics that still allow us to discover this model. 
It clearly shows that even with $100\%$ systematic uncertainties (compared to the statistics uncertainties), we can discover this model across a wide range of its parameter space. 
The mass of the DM could affect this calculation only via the branching ratio of the mediator.
However, Fig.~\ref{fig:BR_L} implies we can safely assume the charged mediator always decays to $\chi \ell$ in this part of the parameter space. Therefore, our analysis is independent of the DM mass in this part of the parameter space.

\subsubsection{Long-lived Particle Region}
\label{subsubsec:llp_L}

We focus on the signal of the charged mediator in the LLP region of its parameter space. 
This is a result of the fact that the neutral mediator of this setup always decays invisibly and, in the LLP region of the parameter space, it leaves no trace in the detector. 
Here, we focus on the dominant decay of the charged mediator in this part of the parameter space, namely to a neutral mediator, which subsequently decays invisibly, and soft leptons -- see~\cref{fig:BR_L}. In this case, the lepton track will be too soft to be detected and we will observe a disappearing track signal. We include events with either two charged mediators or only one in the final state; the latter class includes, \textit{e.g.}, $W^- Z \to \phi^- \phi^0$ processes and make up an $\mathcal{O}(10\%)$ of the total expected charged tracks for $m_\phi \sim \mathcal{O}(1)$~TeV.

This signal has many similarities to the signals studied in Ref.~\cite{Capdevilla:2021fmj} in the context of Wino or Higgsino DM models; nonetheless, the relic abundance calculation in our setup is completely different from the models of Ref.~\cite{Capdevilla:2021fmj}, giving rise to a different relationship between DM mass and mediator's lifetime on the parameter space. 
Ref.~\cite{Capdevilla:2021fmj} includes an extensive GEANT4 simulation, which is required for a proper study of the background. 
In our study, however, we only report the signal yield and refrain from a more complete analysis with the background. 

\begin{figure}
            \centering
            \includegraphics[width=0.9\linewidth]{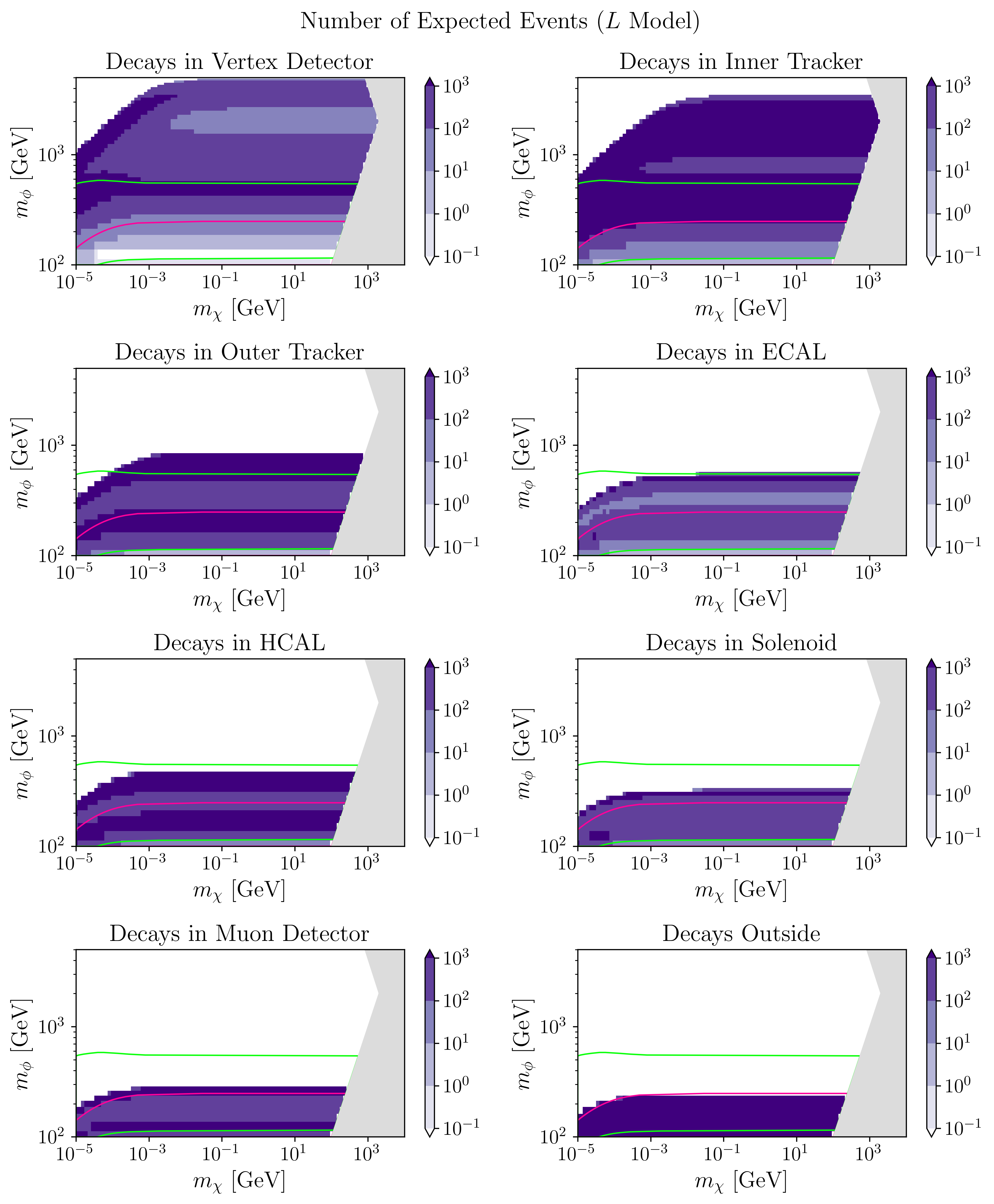}
            \caption{\underline{$L$ model:} Number of expected disappearing tracks and displaced leptons in different barrel regions of the detector, assuming the tentative design of Table~\ref{tab:detector}, as well as number of expected detector-stable charged tracks. This includes all 2-body final states that have at least one charged mediator $\phi^\pm$. 
            The gray region corresponds to $m_\chi > m_\phi$ (lower half) or DM overclosing the universe (upper half) and is not phenomenologically viable. For the entire LLP region of the parameter space, we find $\gtrsim 10^3$ events in some part of the detector. The {\bf\textcolor{xkcdNeonGreen}{green}} and {\bf\textcolor{xkcdNeonPink}{pink}} regions indicate current LHC slepton constraints from Ref.~\cite{ATLAS:2020wjh} and Ref.~\cite{CMS:2024qys}, respectively. }
            \label{fig:llp_result_L_1}
\end{figure}

Using the cross sections in \cref{fig:ellR_xsec}, and the branching ratios in~\cref{fig:BR_L}, we can calculate the event count for this disappearing track signal. 
In~\cref{fig:llp_result_L_1} we show the rate for production of disappearing tracks terminating in different parts of the detector and on a plane of mediator mass versus DM mass.

Owing to the three-body decay via the electroweak interactions, the lifetime of the charged mediator can be very different compared to the $e$ model mediator for the same point on the mass plane. 
The vast majority of the parameter space will have the charged $\phi$ decaying within the detector, leaving a disappearing track signal as mentioned before. Additionally, we see the return of the double-peak feature, manifesting here as two horizontal regions of high event count. 
The inner tracker, in particular, sees a fantastic $>10^3$ event count throughout a majority of the parameter space, with decays from the vertex detector and outer tracker rounding it out. This will likely make the disappearing track search very feasible, and alone could discover this model in most of the LLP region.

\subsection{$u$ model}
\label{subsec:collideru}

In this section we study the collider signatures of the $u$ model. 
The cross section for a pair production of mediators is shown in \cref{fig:ellR_xsec} as well.
Similar to the $e$ model studied above, in this setup we have only one mediator. 
Nonetheless, this mediator is now color-charged, giving rise to jets when it decays, instead of leptons or a single charged track.
As before, in our analysis here we assume the DM has the correct relic abundance today for every point on its parameter space, which can be used to find its coupling to mediators, see \cref{fig:lambda_lifetime_u}.

The mediator in this setup has only one decay channel, namely to a DM candidate and SM quark of the same charge. 
As a result, unlike the $L$ model above, its decay products are always detectable in both the prompt and the LLP regimes.

\subsubsection{Prompt Region}
\label{subsubsec:prompt_u}

In this part of the parameter space, after its production, the heavy mediator promptly decays to a SM quark and missing energy. 
There are many standard techniques for searching for such resonances that decay to jet+MET. 
Using the cross section from \cref{fig:ellR_xsec} we can calculate the production rate of these mediators and then use various kinematic cuts to look for them in the data.
In our analysis we will use the momentum of the quark from the $\phi_u$ mediator decay as the proxy for the jet momentum.

The background topologies are the same as the previous models, but now with SM quarks, instead of leptons, in the final state, see \cref{fig:bkgd_diagrams}. Cuts on similar observables as in the lepton-portal can subtract the background from the signal in this model as well. 
The optimal cuts on these kinematic variables for two different benchmark masses $m_\phi = 1.6$~TeV and $m_\phi = 4.2$~TeV, are respectively
\begin{align}
    \label{eq:SR_u_def}
    \mathrm{SR}_1^q(m_\phi=1.6~\rm{TeV})&: \left\lbrace m_{jj} \geqslant 3600~\mathrm{GeV}, \theta_{jj} \geqslant \frac{7\pi}{9},|\eta_{j}| \leqslant 0.5, M_{T2} \geqslant 400~\mathrm{GeV}, E_{j} \geqslant 900~\mathrm{GeV}  \right\rbrace , \\
    \mathrm{SR}_2^q(m_\phi=4.2~\rm{TeV})&: \left\lbrace  m_{jj} \geqslant 1500~\mathrm{GeV}, \theta_{jj} \geqslant \frac{\pi}{4},|\eta_{j}| \leqslant 0.7, M_{T2} \geqslant 1600~\mathrm{GeV}, E_{j} \geqslant 1600~\mathrm{GeV}  \right\rbrace.
    \nonumber 
\end{align}
We choose to optimize the signal regions for higher mediator masses (compared to the lepton--portal models) as the current LHC bounds on color-charged particles are stronger at $m_\phi\gtrsim 1.6$ TeV~\cite{CMS:2019zmd,CMS:2019ybf,ATLAS:2020syg}.\footnote{Our mediator decays to five flavors of RH quarks; we scale the theory cross sections accordingly to calculate the bounds from Refs.~\cite{CMS:2019zmd,CMS:2019ybf,ATLAS:2020syg} on our model.  } 

\begin{figure}
            \centering
            \resizebox{\columnwidth}{!}{
            \includegraphics[width=0.6\linewidth]{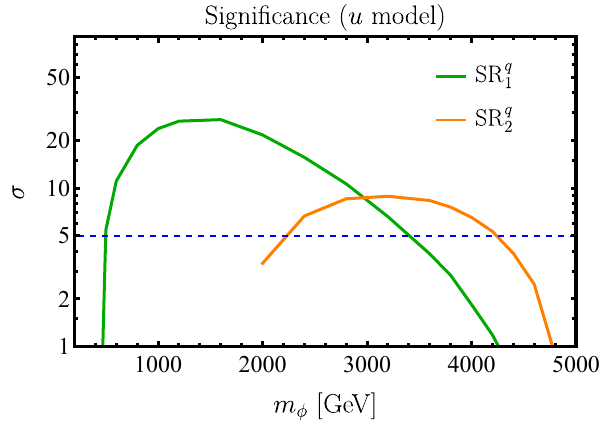}
            \hspace{0.1in}
            \includegraphics[width=0.62\linewidth]{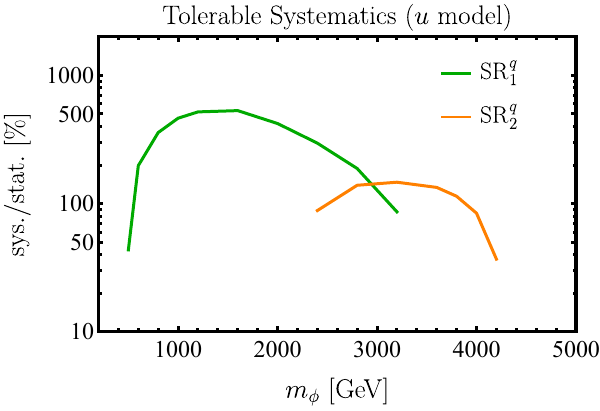}
            }
            \caption{\underline{$u$ model:} The best achievable discovery significance with zero systematic uncertainty (\textbf{left}) and the tolerable uncertainty that still allows a $5\sigma$ discovery of the mediator (\textbf{right}) for the signal regions defined in Eq.~\eqref{eq:SR_u_def}. The dashed blue line on the left plot marks the $5\sigma$ discovery threshold. We find that even with systematic uncertainties as large as the statistical ones the model is discoverable in a future muon collider for mediator masses up to $m_\phi \lesssim 4.2$~TeV. }
            \label{fig:prompt_result_u}
\end{figure}

The production diagram for this model is similar to the lepton--portal model, see \cref{tab:diagrams}. The production rate is enhanced by a color factor and is larger than the lepton--portal model's rate. As a result, we expect a higher reach in the parameter space of this models. 
We show the discovery reach of a 10~TeV muon collider for this model in \cref{fig:prompt_result_u}. 
We find that using the signal regions from Eq.~\eqref{eq:SR_u_def} allows us to discover this model for $m_\phi \lesssim 4.2$~TeV, which is a large improvement on the current bounds of $m_\phi \gtrsim 1.6$~TeV from LHC \cite{CMS:2019zmd,CMS:2019ybf,ATLAS:2020syg}.

We also report the tolerable systematics that still allows different signal regions to discover the model for different mediator masses in the same figure. 
Similar to the previous models, we find that even with systematic uncertainties as big as the statistical ones, our rudimentary analysis can discover this model for a vast range of kinematically allowed mass ranges.
As these fermion portals constitute well-motivated targets for future collider searches, the tolerable systematics in this figure can serve as a motivated target in future designs of the detector for a muon collider machine.

\subsubsection{Long-lived Particle Region}
\label{subsubsec:llp_u}

We now turn our focus to the signal of the charged mediator in the LLP regime of its parameter space, see~\cref{fig:lambda_lifetime_u}. Since our mediator now has color charge, its LLP signature will be $R$-hadrons \cite{Farrar:1978xj,Arkani-Hamed:2004ymt}.\footnote{In what follows, we will refer to these LLP objects as $R$-hadrons, even though strictly speaking we are not studying a supersymmetric theory.} Because the mass of the mediator is much larger than the quarks it will bind with, we assume the mass and lifetime of the resulting $R$-hadrons will essentially be unchanged from those of the mediator itself~\cite{ATLAS:2019duq}. 
Before going into the details of our analysis and results, it is useful to review what happens to an $R$-hadron as it passes through different parts of the detector.

A large fraction of $R$-hadrons produced at colliders would be charged, allowing them to give rise to an anomalous ionization rate, $dE/dx$, as they pass through the pixel detector. 
They appear as charged tracks in the tracker system and can be easily detected. 
It should, however, be noted that through rearrangement processes with the detector material, these hadrons can change charge (and spin) and give rise to intriguing \textit{sporadic} tracks and a unique ionization rate \cite{Arkani-Hamed:2004ymt,Arvanitaki:2005nq}. 
That said, given the material and the density of the typical tracker systems, the probability of a charge flip in this detector component is quite low, \textit{e.g.} at the percent level at LHC \cite{ATLAS:2019duq}. 

If a neutral $R$-hadron decays in the tracker system, it gives rise to a detectable displaced jet.
On the other hand, if the decaying $R$-hadron is charged, the signal consists of a charged track that ends at a displaced vertex in the tracker system. 
In either case, the displaced vertex can be detected and serve as a signal of the $R$-hadron.

Unlike in the tracker system, the $R$-hadron does not leave a detectable track as it passes through the calorimeters. 
The amount of deposited energy via anomalous ionization is very low in the calorimeters, making the measurement of $dE/dx$ signal virtually hopeless in this detector component, similar to LHC.
On the other hand, given the higher density of the material in the calorimeter, the $R$-hadron can go through multiple rearrangement processes that changes its charge (and spin) multiple times.
Unless they decay in the calorimeters, it is safe to assume they leave no detectable signal in there as they pass through. 
A displaced jet from their decay can be easily detected though.

When passing through the muon system, a charged $R$-hadron again appears similar to a charged track. 
While the LHC does not use the $dE/dx$ information in the muon system, the $R$-hadron does deposit some small energy into the system as it passes through, which could in principle be used as well. 
Depending on the design of the muon system, the track can go through rearrangement processes that change its charge to give rise to sporadic tracks. 
For instance, such processes would not occur frequently in the hollowed ATLAS muon system, whereas they can potentially take place in CMS denser muon system. 
An $R$-hadron decaying in the muon system can give rise to a detectable displaced jet as well.

Finally, if they are long-lived enough to be detector-stable, they traverse through all detector layers, leaving detectable (and potentially sporadic) tracks in the tracker and the muon system, while leaving no trace in the calorimeter.

\begin{figure}
            \centering
            \includegraphics[width=0.9\linewidth]{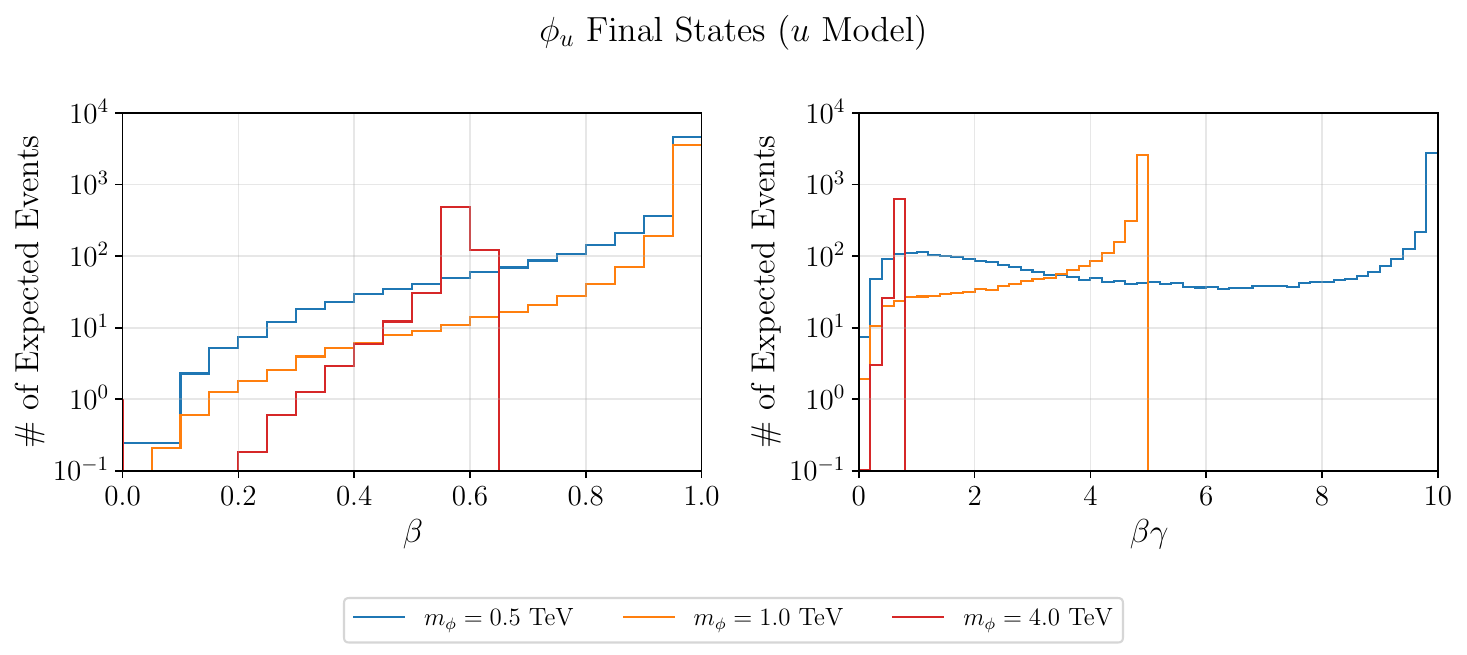}
            \includegraphics[width=0.9\linewidth]{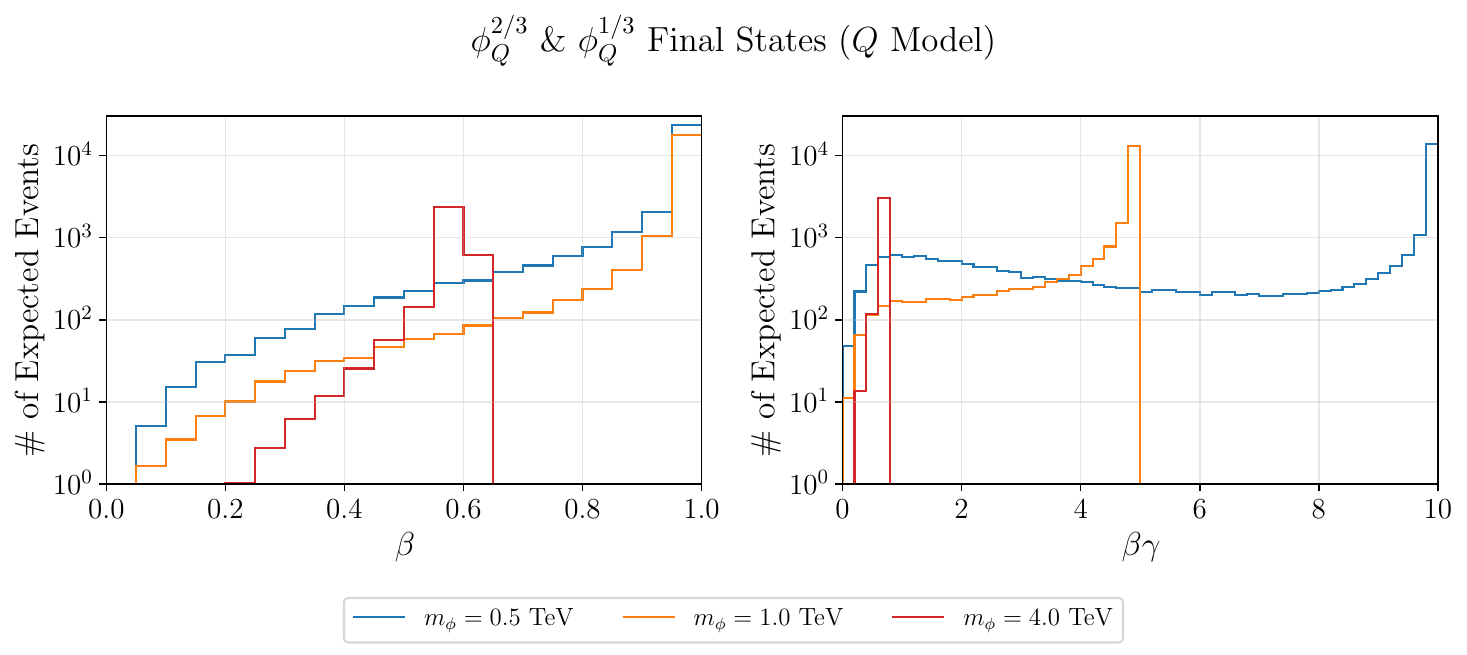}
            \caption{The $\beta$ (\textbf{left}) and $\beta \gamma$ (\textbf{right}) distribution of events for three different mediator masses ({\bf\textcolor{matBlue}{blue}} = 500 GeV, {\bf\textcolor{matOrange}{orange}} = 1 TeV, {\bf\textcolor{matRed}{red}} = 4 TeV) in the $u$ model (\textbf{top}) and the $Q$ model (\textbf{bottom}). These distributions are used in determining various signals of LLP signals, \textit{e.g.} the time-of-flight and the $R$-hadron stopping signal. The peak at large values of $\beta$ and $\beta \gamma$ arises from muon annihilation channel.   }
            \label{fig:beta_dist}
\end{figure}

A useful quantity in the study of the $R$-hadron signals is the distribution of their velocity after production. 
This distribution is relevant for various LLP signals such as time-of-flight measurements, $dE/dx$, or $R$-hadrons stoppage signal \cite{Arvanitaki:2005nq}. 
For both the $u$ model and the $Q$ model (next section), this distribution is included in \cref{fig:beta_dist}. 
We can see that a vast majority of the events have a very large velocity $\beta$. 
The small fraction of events with low velocity suppresses the stopping $R$-hadron signature. 
At LHC this stopping $R$-hadron signature is less efficient than other search strategies such as $dE/dx$ or displaced jet searches \cite{ATL-PHYS-PUB-2024-014}, as only the very slow-moving ones can be stopped. 
We expect the fraction of $R$-hadrons that move slow enough decreases in a high-energy muon collider, making them an even less efficient signal. As a result, we do not study this signal in our search for $R$-hadrons.

\begin{figure}
            \centering
            \includegraphics[width=0.9\linewidth]{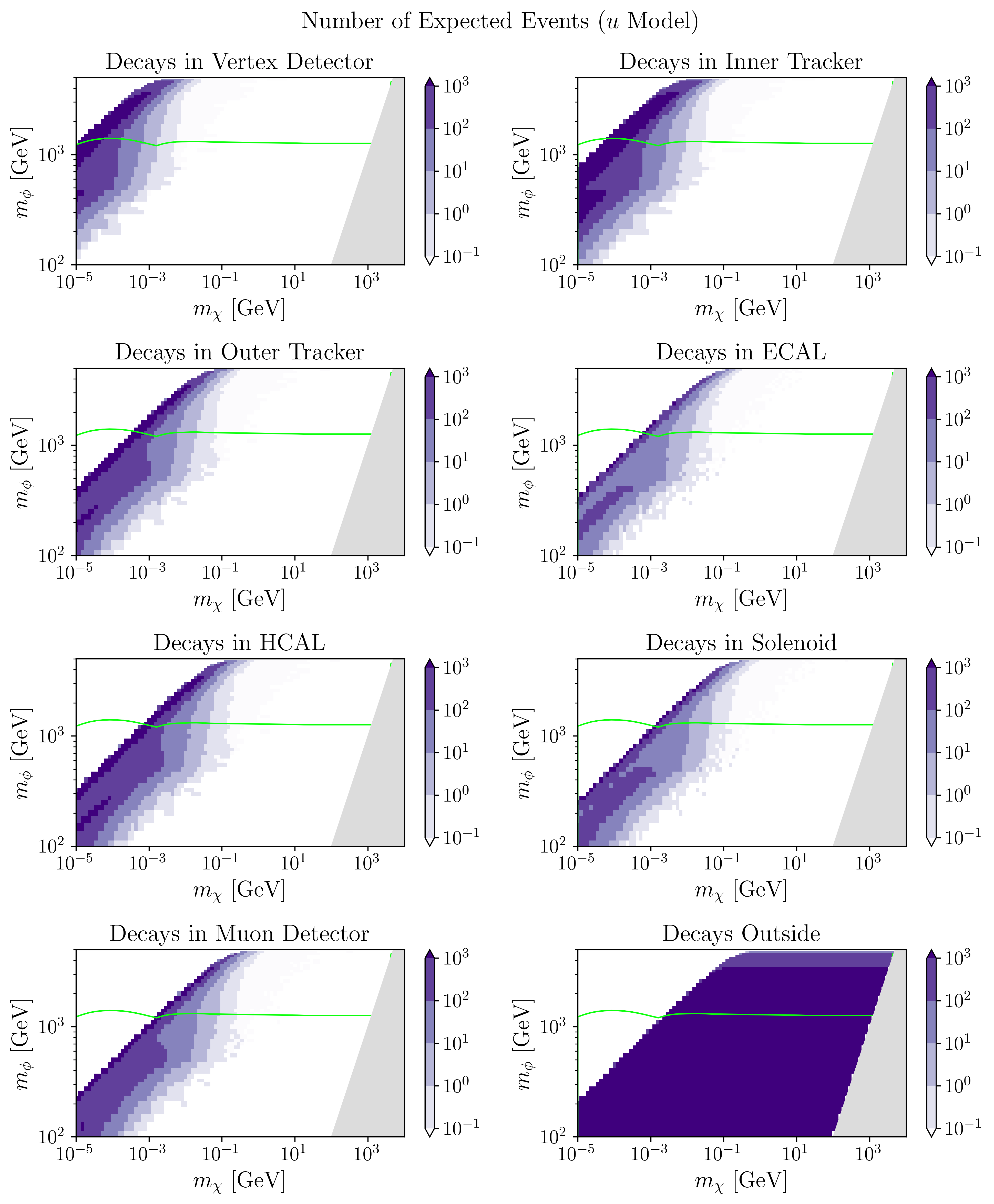}
            \caption{\underline{$u$ model:} Number of expected displaced vertices in different barrel regions of the detector assuming the tentative design of Table~\ref{tab:detector}, as well as number of expected detector-stable $R$-hadrons. 
            The gray region corresponds to $m_\chi > m_\phi$ (lower half) or DM overclosing the universe (upper half) and is not phenomenologically viable. For the entire LLP region of the parameter space, we find $\sim 10^3$ events in some part of the detector. The {\bf\textcolor{xkcdNeonGreen}{green}} regions indicate current LHC squark constraints from Ref.~\cite{ATL-PHYS-PUB-2024-014}.  }
            \label{fig:llp_result_u}
\end{figure}

While studying properties of the $R$-hadron at a muon collider is an intriguing research direction, for the level of the analysis in this work we focus on the displaced jet signal, in line with the search proposal for the lepton-portal models in the previous sections.
In~\cref{fig:llp_result_u} we report the average number of decay events within each detector system as in the lepton LLP sections. 
This figure shows the average number of expected $R$-hadron decays in the different detector systems. 
The qualitative shape of the high event count parts of the parameter space is very similar to the $e$ model; this arises from the fact that in both these models, the mediator only has one decay channel (to DM and a SM fermion) with comparable lifetimes.

Similar to the $e$ model, for most of the parameter space, the mediator decays outside the detector. 
As discussed above, such long-lived $R$-hadron can give rise to an exotic sporadic track in the tracker and the muon system, and leave no trace in the calorimeter.
If instead the mediator does decay in a detector component, it gives rise to a displaced jet signature. 
Similar to the previous models, we find that for every point in the LLP region of the parameter space, there will be some component of the detector with at least $\sim 10^3$ events.

\subsection{$Q$ model}
\label{subsec:colliderQ}

Similar to the previous models, in our analysis we consider the production of a pair of mediators and their subsequent decay. 
Unlike previous models, in this model we have two mediators that both decay visibly; as a result, we will have a larger signal rate that allows us to discover this model up to higher mediator masses.
The total cross section for this pair production is shown in \cref{fig:ellR_xsec}. 
We will find that in the prompt region, our analysis significantly improves the current LHC bounds of $m_\phi \gtrsim 1.6$~TeV \cite{CMS:2019zmd,CMS:2019ybf,ATLAS:2020syg}, and in the LLP region we find a large event count from $R$-hadrons in different detector components.

\subsubsection{Prompt Region}
\label{subsubsec:prompt_Q}

The background for the $Q$ model signal is the same as the $u$ model signal discussed in the previous section. 
Since the signal final state is similar as well, we use the same signal regions as in the $u$ model, see Eq.~\eqref{eq:SR_u_def}, for our search here as well.

\begin{figure}
            \centering
            \resizebox{\columnwidth}{!}{
            \includegraphics[width=0.6\linewidth]{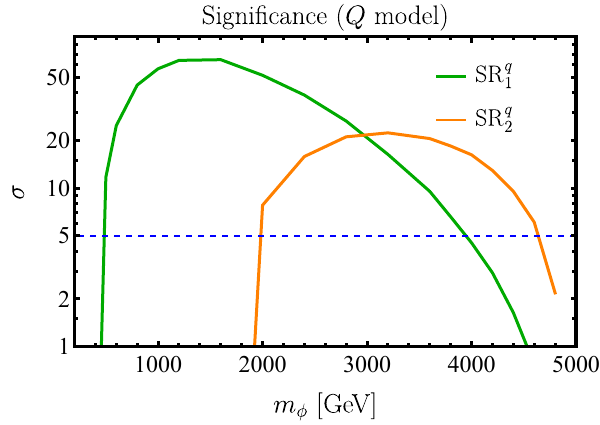}
            \hspace{0.1in}
            \includegraphics[width=0.62\linewidth]{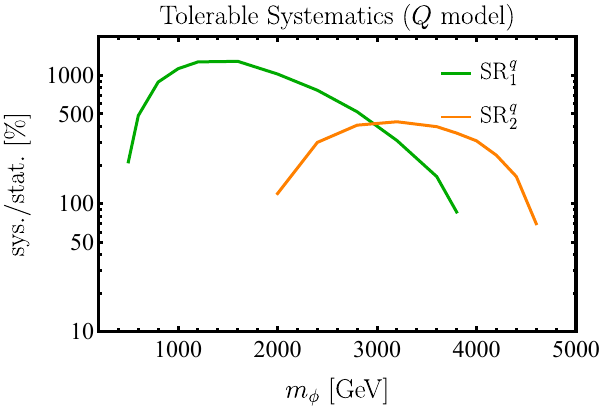}
            }
            \caption{\underline{$Q$ model:} The best achievable discovery significance with zero systematic uncertainty (\textbf{left}) and the tolerable uncertainty that still allows a $5\sigma$ discovery of the mediator (\textbf{right}) for the signal regions defined in Eq.~\eqref{eq:SR_u_def}. The dashed blue line on the left plot marks the $5\sigma$ discovery threshold. We find that even with systematic uncertainties as large as the statistical ones the model is discoverable in a future muon collider for mediator masses up to $m_\phi \lesssim 4.7$~TeV. }
            \label{fig:prompt_result_Q}
\end{figure}

The attainable discovery significance with these signal regions is shown in \cref{fig:prompt_result_Q}. 
We find that these two signal regions allow us to discover this model up to mediator masses of $m_\phi \sim 4.7$~TeV (almost the entire kinematically accessible mass range).
This better reach (compared to the $u$ model) can be attributed to the higher production rate for the mediators and the fact that we have two mediators that decay visibly (to a jet and MET) in this model. 
Unsurprisingly, we find that the tolerable systematics are higher in this model as well, as shown in the right panel in \cref{fig:prompt_result_Q}.

\subsubsection{Long-Lived Particle Region}
\label{subsubsec:llp_Q}

In the LLP region, the lighter mediator behaves similar to the $u$ model scenario: it can bind with SM quarks to give rise to an $R$-hadron signal; similar to the $u$ model, this light mediator dominantly decays to DM $\chi$ and a SM quark, giving rise to a displaced jet signature. 

The signal from a heavy mediator decay is more complicated. 
For almost the entire LLP part of the parameter space, the heavier mediator dominantly decays to the lighter one and very soft electron and neutrino that will go undetected. 
The lighter mediator then decays later (since it has only one decay channel and thus a longer lifetime than the heavier mediator), giving rise to a displaced jet signal as discussed above. 
As a result, for the heavy mediators we consider this full decay chain through the lighter mediator to an eventual displaced jet signal. We assume the soft SM leptons from the initial decay of the heavy mediator are soft enough to go undetected and do not change the momentum of the mediator before and after the decay.

\begin{figure}
            \centering
            \includegraphics[width=0.9\linewidth]{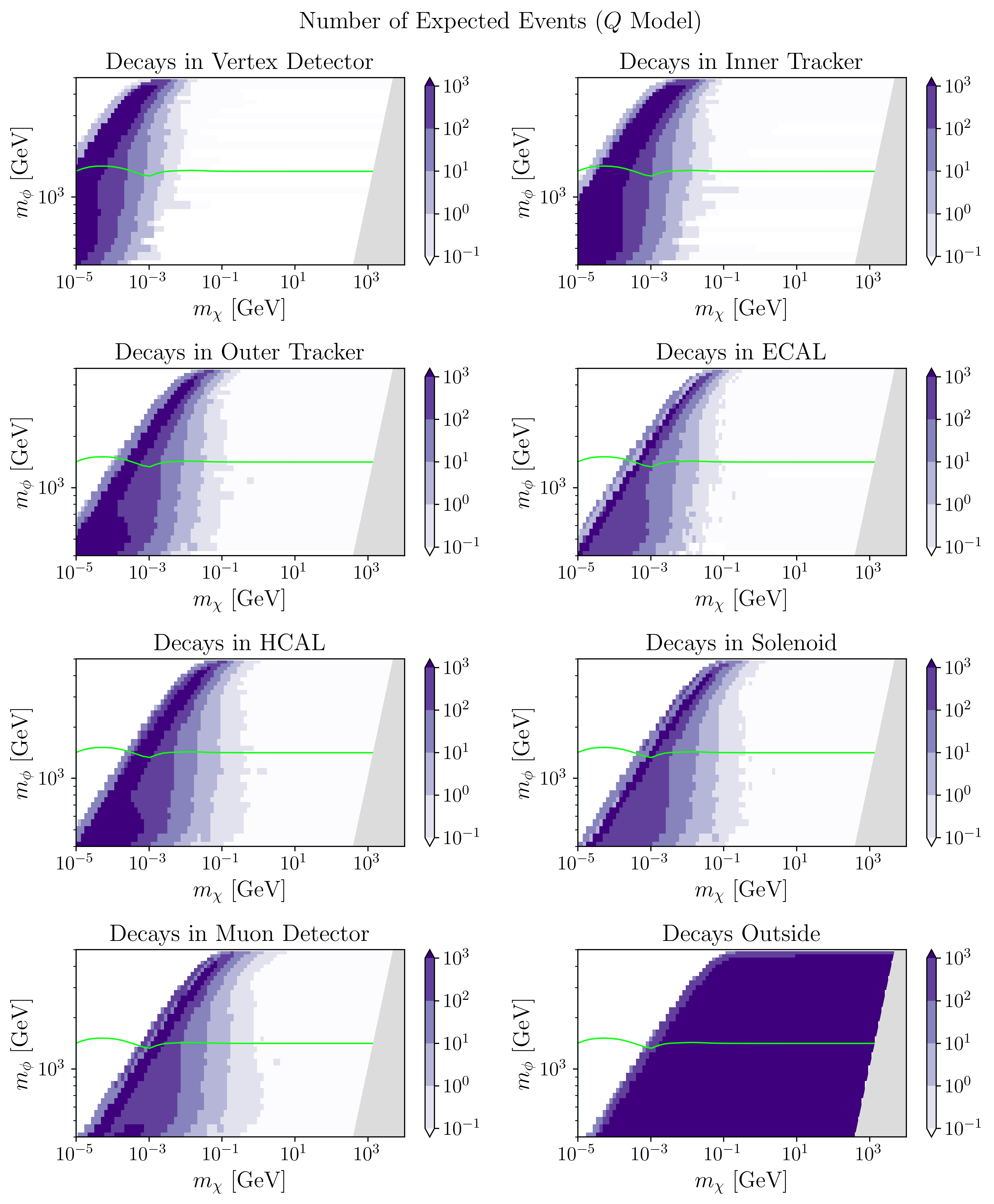}
            \caption{\underline{$Q$ model:} Number of expected displaced vertices in different barrel regions of the detector assuming the tentative design of Table~\ref{tab:detector}, as well as number of expected detector-stable charged tracks. The total count includes the direct decay of light mediators, $\phi^{1/3}$, as well as the decay chain of the heavy mediator through a light mediator (heavy mediator $\rightarrow$ light mediator $\rightarrow$ SM quarks). 
            The gray region corresponds to $m_\chi > m_\phi$ (lower half) or DM overclosing the universe (upper half) and is not phenomenologically viable. For the entire LLP region of the parameter space, we find $\sim 10^3$ events in some part of the detector. The {\bf\textcolor{xkcdNeonGreen}{green}} regions indicate current LHC squark constraints from Ref.~\cite{ATL-PHYS-PUB-2024-014}.}
            \label{fig:llp_result_Q}
\end{figure}

In \cref{fig:llp_result_Q}, we report the average number of displaced jets appearing in different components of the detector. This includes the jets from direct production and decays of the lighter mediator, as well as from a heavy mediator that decays to DM and a jet via an intermediate on-shell light mediator.

We find some features of the $L$ model (\cref{fig:llp_result_L_1}) and the $u$ model (\cref{fig:llp_result_u}) in the these figures. 
Compared to the $u$ model signals, regions of high event counts are stretched out to higher $m_\chi$ (or $\tau_\phi$) values; this can be attributed to the two-staged decay chain above (heavy mediator $\rightarrow$ light mediator $\rightarrow$ SM quarks). 
The vast majority of points in the LLP parameter space give rise to a high number of $R$-hadron decays ($\gtrsim 10^3$) either in the tracker system or outside the detector, respectively, similar to the $L$ model or the $u$ model.

\section{Discussion and Conclusions}
\label{sec:conclusion}

In this work, we study four different models of fermion-portal DM in the freeze-in regime and their signatures at a 10 TeV muon collider: 
the RH lepton-portal model ($e$ model), the LH lepton-portal model ($L$ model), the RH quark-portal model ($u$ model) and the LH quark-portal model ($Q$ model). For the $e$ model, we expand on previous work~\cite{Asadi:2023csb} to include muon component PDFs in the collider study.  
We find some results that apply to all four models. Due to the interplay between the direct freeze-in and the superWIMP mechanisms, there is an upper bound of $\mathcal{O}$(1-10)~TeV on the DM mass in all these models, which allows for a vast portion of the viable parameter space of these models to be studied at a future 10 TeV muon collider. 
Furthermore, the two different production channels, \textit{i.e.} muon annihilation or VBF, give rise to drastically different distributions that manifest themselves in a double-peak behavior in various kinematics distributions, and provide valuable handles to separate signal from background.

Each model could give rise to either prompt or LLP signals.
In the prompt regime, we proposed simple cut-and-count searches. 
The main signal in these models arises from the pair production and subsequent decays of the mediator particles to missing energy and a pair of leptons (for the lepton-portal models) or jets (for the quark-portal models). 
In the LLP region we used the number of charged tracks (lepton-portal) or $R$-hadrons (quark-portal) that either decay in the detector or escape.

For the first time in the literature, we also concocted a  pipeline for including the non-trivial muon parton PDF in a muon beam in the event simulation routine. 
To include the effect of the non-trivial muon component PDF, we decompose it into two pieces: (1) its previously calculated PDF for $x \leqslant 1 - \epsilon$ values, and (2) a $\delta (x-1)$ piece whose normalization is determined by fixing the net muon number of the beam and the ratio of different muon polarizations therein. 
While the total cross section is not significantly affected, we found that this non-trivial muon PDF distorts the signal event distributions enough so that they have an indispensable effect in the final reach of a muon collider in the parameter space of fermion-portal models.
In particular, we showed that previously proposed search strategies~\cite{Asadi:2023csb} are significantly less efficient and new cuts have to be deployed to discover these models.

For these fermion-portal models, cuts on $m_{\ell\ell,jj}$, the opening angle between the pair of leptons or jets ($\theta_{\ell\ell,jj}$), the energy of each lepton or jet ($E_{\ell,j}$), $M_{T2}$, and the pseudorapidity ($\eta$) of visible final states can efficiently subtract the background. 
We found that when systematic uncertainties are neglected, the discovery reach of a 10 TeV muon collider with 10 ab$^{-1}$ data spans from 3.3 to 4.7~TeV in the mediator mass for different portal models, see \cref{sec:results} for more accurate numbers. 
This reach does not diminish significantly with systematics uncertainties that are comparable to the statistical uncertainties.

In the LLP regime, these models can give rise to heavy stable charged particles, displaced leptons or vertices, and disappearing charged tracks or $R$-hadrons that eventually decay and give rise to a displaced jet. 
We found a large event count ($\sim 10^3$) in at least one detector component for the entire LLP region of each model's parameter space. 

In the RH lepton-portal model ($e$ model), the mediator can only decay via the freeze-in coupling and can become detector stable; in these models the heavy stable charged particle signal is the primary signal for most of the LLP region. 
In the RH quark-portal model ($u$ model), the signal will be an $R$-hadron that, if it decays inside the detector, gives rise to a displaced jet signal. 
Nonetheless, in most of its LLP parameter space, the $R$-hadron decays outside the detector, giving rise to a track in the tracker and the muon system. 
As the $R$-hadron can change charge via rearrangement processes with the detector, these would be sporadic tracks, a smoking gun signature of $R$-hadrons.

In the LH lepton-portal model ($L$ model), the heavy mediator decays dominantly through electroweak interactions to the lighter mediator. As a result, these mediators are not as long-lived as their RH counterparts. 
We find that for these models the main signal will be displaced leptons in the tracker system. 
In the LH quark-portal model ($Q$ model) the heavy mediator decays mostly to the lighter mediator, both of which appear as $R$-hadrons. 
In this model for most of the LLP parameter space the $R$-hadron is detector stable, similar to the $u$ model $R$-hadrons.

Our findings suggest potential refinements to the MUSIC detector design that could enhance its sensitivity to fermion-portal dark matter, particularly within the LLP regime.
The main obstacle in many LLP searches 
is an efficient layer~1 trigger on their signals which has motivated both theory and experimental studies, see \textit{e.g.}, Refs.~\cite{Gershtein:2019dhy,Gershtein:2020mwi} for recent works on new trigger proposals at CMS and their applicability to LLP searches. 
In particular, CMS has recently considered double-layered sensors in their tracker system that gives rise to a \textit{stub} (arising from a double hit), which in turn provides a handle on the $p_T$ of the track~\cite{CMS:2017lum,Zabi:2020gjd}. 
This enables a trigger on isolated high $p_T$ tracks and could be the main trigger for the LLP signals of all fermion portals studied here.

A suitably designed muon system can help in detecting heavy stable charged tracks, which are the main signal of three (out of four) models studied here across most of their available parameter space. 
Thus, our signals could potentially inform the on-going research on the design of the muon system in MUSIC, \textit{e.g.} see Refs.~\cite{MuonCollider:2022ded,Aime:2022mpz}. 
Specifically, a slow muon trigger can discern these tracks from the large SM background.\footnote{We thank Laura Jeanty for pointing out this possibility.} 
If the possibility of measuring $dE/dx$ in the muon system exists, it can be used for tagging these signals with a high efficacy as well.

These improvements are also crucial for reconstructing the sporadic track signal of our models in the layered muon system. 
Reconstruction of momentum and charge of the stable charged tracks in a layered muon system is already a complicated task. 
For the sporadic track signal discussed in the text, the change in the charge of the track between each layer of detector can further complicate this task significantly.
It should be noted that while the sporadic tracks in the muon system are a smoking gun signatures of detector-stable $R$-hadrons, they are not necessary for their discovery. 

The choice of whether to prioritize the detection of sporadic tracks in the muon system can significantly influence its design. For example, the CMS muon system, with dense absorber material between detector layers, increases the likelihood of $R$-hadron charge flipping, resulting in sporadic tracks. In contrast, the hollowed design of the ATLAS muon system eliminates this effect. As we consider the design of the MUSIC muon detector, it is crucial to account for the potential implications of such sporadic signals and the complex challenges associated with their reconstruction. Addressing these considerations will ensure an optimized and robust detector design.

\section*{Acknowledgments}

We are particularly in debt to Laura Jeanty for many illuminating discussions on LLP signatures, especially of $R$-hadrons. 
We also thank Austin Batz, Spencer Chang, and Ben Lillard for helpful discussions. 
We thank Sokratis Trifinopoulos for discussions on  LePDF.
The work of P.A. is supported by the U.S. Department of Energy under grant number DE-SC0011640. The work of SH is supported by the NSF grant PHY-2309456. T.-T.Y. and A.R. are supported in part by NSF CAREER grant PHY-1944826. T.-T.Y. thanks the Università Degli Studi di Padova for their hospitality and support where part of this work was completed. 

\appendix

\section{Mass Splitting}
\label{appx:masssplitting}
Two of the fermion-portal models, the $L$  and $Q$ models, come with a new scalar doublet. 
At tree-level, the two components of the new doublet have the same mass $M$. 
The mass difference, which is induced by loops of SM gauge bosons between the two components of the new doublet (with hypercharge $Y$) and component electric charges $Q$ and $Q'$, is given by~\cite{Cirelli:2005uq}
\begin{multline}
    M_Q - M_{Q'} = \frac{\alpha_2 M}{4 \pi} \bigg\{ (Q^2-Q'^2) s_W^2 f\left(\frac{m_Z}{M}\right) \\ + (Q-Q')(Q+Q'-2Y) \Big[ f\left(\frac{m_W}{M}\right) - f\left(\frac{m_Z}{M}\right) \Big] \bigg\}\, ,
\end{multline}
where $\alpha_2 = \alpha_{\textrm{EM}}/s_W^2$, and
\begin{align}
    f(r) = -\frac{r}{4} \left[ 2 r^3 \ln r + (r^2-4)^{3/2} \ln A(r) \right]\, ,
\end{align}
where
\begin{align}
    A(r) = \frac{1}{2} (r^2-2-r \sqrt{r^2-4})\, .
\end{align}
For the $L$ model, $Y=-1/2$, $Q=-1$, and $Q'=0$ leading to
\begin{align}
    M_{\phi^-} - M_{\phi^0} = \frac{\alpha_{\textrm{EM}} M}{4 \pi} f\left(\frac{m_Z}{M}\right) \xrightarrow[]{M \to \infty} 0.333 \textrm{ GeV}\, .
\end{align}
For the $Q$ model, $Y=1/6$, $Q_u = 2/3$, $Q_d = -1/3$ and so
\begin{align}
    M_{\phi_u} - M_{\phi_d} = \frac{\alpha_{\textrm{EM}} M}{12 \pi} f\left(\frac{m_Z}{M}\right) \xrightarrow[]{M \to \infty} 0.111 \textrm{ GeV}\, ,
\end{align}
or exactly $1/3$ the mass splitting for the $L$ model.

\end{spacing}
\newpage
\bibliographystyle{utphys}
\bibliography{bib}
\end{document}